\documentclass[prb,showpacs,twocolumn,superscriptaddress,citeautoscript,twoside]{revtex4}

\usepackage{amsmath,graphicx,latexsym,bm}

\begin{document}

\title{First-principles modeling of ferroelectric capacitors via 
constrained-${\bf D}$ calculations}

\author{Massimiliano Stengel}
\affiliation{CECAM - Centre Europ\'een de Calcul Atomique et Mol\'eculaire,
Station 13, Bat. PPH, 1015 Lausanne, Switzerland}
\author{David Vanderbilt}
\affiliation{Department of Physics and Astronomy, Rutgers University, Piscataway,
             New Jersey 08854-8019, USA}
\author{Nicola A. Spaldin}
\affiliation{Materials Department, University of California, Santa Barbara,
             CA 93106-5050, USA}

\date{\today}

\begin{abstract}
First-principles modeling of ferroelectric capacitors presents several
technical challenges, due to the coexistence of metallic electrodes,
long-range electrostatic forces and short-range interface chemistry.
Here we show how these aspects can be efficiently and accurately
rationalized by using a finite-field density-functional theory formalism
in which the fundamental electrical variable is the displacement field 
$\mathbf{D}$.
By performing calculations on model Pt/BaTiO$_3$/Pt and Au/BaZrO$_3$/Au
capacitors we demonstrate how the interface-specific and bulk-specific 
properties can be identified and rigorously separated.
Then, we show how the electrical properties of capacitors of 
arbitrary thickness and geometry (symmetric or asymmetric) can be 
readily reconstructed by using such information.
Finally, we show how useful observables such as
polarization and dielectric, piezoelectric and electrostrictive coefficients
are easily evaluated as a byproduct of the above procedure.
We apply this methodology to elucidate the relationship between chemical 
bonding, Schottky barriers and ferroelectric polarization at  
simple-metal/oxide interfaces. We find that BO$_2$-electrode interfaces behave 
analogously to a layer of linear dielectric put in series with a bulk-like
perovskite film, while a significant non-linear effect occurs at 
AO-electrode interfaces.
\end{abstract}


\maketitle

\marginparwidth 2.7in
\marginparsep 0.5in
\def\msm#1{\marginpar{\small MS: #1}}
\def\dvm#1{\marginpar{\small DV: #1}}
\def\nsm#1{\marginpar{\small NS: #1}}
\def\scr{\scriptsize}
  \def\msm#1{}
  \def\dvm#1{}
  \def\nsm#1{}

\def\Ef{E_{\rm f}}
\def\e{\bar\varepsilon}

\section{Introduction}

Capacitors based on ferroelectric perovskites hold promise for
substantial advances in nanoelectronics, with potential applications in
non-volatile random-access memories and high-permittivity gate
dielectrics~\cite{Scott:2007}.
Thinner devices, which are mandatory for optimal efficiency and speed,
are strongly influenced by the electrical and mechanical boundary
conditions imposed by the interface~\cite{Dawber:2005}.
While there has been significant progress in the understanding of
strain effects~\cite{Rabe_strain}, the electrostatics of metal-ferroelectric 
interfaces still remains a challenge, and is widely recognized as 
a central issue in the scaling of ferroelectric devices.

Interface electrostatics is generally modeled, within Landau-Ginzburg
theories, by a hypothetical thin layer of standard dielectric (``dead layer'')
interposed between an ideal electrode and the active, bulk-like ferroelectric
film.
The dielectric dead layer is arranged in series with the film, and therefore
the small interfacial capacitance associated with it tends to suppress
the polarization of the film via a depolarizing field~\cite{Junquera_review:2008}.
It was postulated a long time ago~\cite{Mead:1961} that, even in the absence of 
an extrinsic interfacial layer, a small interfacial capacitance can originate from
the finite penetration length of the electric field in a real electrode.
The imperfect-screening model and the dead-layer model are
mathematically equivalent and lead to the same consequences, 
regardless of the microscopic nature of the effect~\cite{Bratkovsky_review}.

Owing to the complex structure and chemistry of a realistic interface,
however, it is difficult to infer the magnitude of this interfacial
capacitance based on macroscopic considerations.
Moreover, the usual assumption that the capacitance (or equivalently,
the effective screening length) is \emph{constant}
as a function of the ferroelectric displacement might not be justified
in some cases.
For example, it was shown very recently by means of first-principles
calculations that chemical bonding across the junction profoundly
influences the ferroelectric properties of the device~\cite{nature_mat}.
This is likely to introduce nonlinearities in the electrical response of
the interface that are neglected within most phenomenological
approaches.
In order to achieve a quantitative model of the electrode/ferroelectric
interface there is therefore the clear need for a theory that complements
Landau free-energy expansions with a microscopically reliable description
of local chemistry and electrostatics.

A strategy for modeling the ferroelectric behavior of symmetric and
asymmetric capacitors that combines Landau theory with first-principles
calculations was recently proposed by Gerra \emph{et al.}~\cite{Gerra:2007}
Their strategy has the advantage of exploiting the power of the
\emph{ab-initio} approach to gain quantitative insight into the
coefficients that describe the behavior of the interface.
In particular, the interface enters the free energy through two distinct
quadratic terms, a depolarizing effect which provides a uniform
electric field and is the main contribution, and a short-range
chemical bonding effect which provides a much smaller correction.
These coefficients are then input into a standard Landau
free-energy expansion and used
to predict the behavior of devices of macroscopic thicknesses,
which are not directly tractable from first principles.
This model was shown to describe the SrRuO$_3$/BaTiO$_3$/SrRuO$_3$
system quite accurately. The results were consistent with the seminal
work of Junquera and Ghosez~\cite{Junquera/Ghosez:2003}, who demonstrated
how the main impact of the electrodes is embodied in the
depolarizing field in an otherwise bulk-like BaTiO$_3$ film.

Some authors, however, have questioned the generality of such an
assumption, postulating that in some cases the electrodes can
have a much more profound impact on the ferroelectric film.
The authors of Ref.~~\onlinecite{Duan/Tsymbal:2006},
for example, claimed that
SrRuO$_3$ electrodes can destroy the polar soft mode of
ferroelectric KNbO$_3$ films, producing a head-to-head domain wall
a few unit cells from the interface.
Furthermore, in Ref.~~\onlinecite{Nardelli}, Pt electrodes were found
to induce a ``ferrielectric'' dipole pattern in the whole volume
of a BaTiO$_3$ film.
Such effects, which are nonlocal in nature, cannot be described by
the simple models of Refs.~\onlinecite{Gerra:2007} and 
\onlinecite{Junquera/Ghosez:2003}.
To account for (and clarify the nature of) these ``exceptions'', it
would be very desirable to have a rigorous methodological framework
that treats the electrical properties of a given capacitor heterostructure
fully from first-principles, without any {\it a priori} assumptions.

Such a methodological framework was recently developed for the case 
of purely insulating perovskite superlattices.
By performing the calculations at a fixed value of the electric displacement
field $D$, Wu \emph{et al.}~\cite{Xifan_sl} were able to separate the long-range
electrostatic interactions between layers from the short-ranged compositional
dependence.
Based on this separation, the electrical properties of a given layer were
shown to depend on the chemical nature of a small number of first and second
neighbors only.
This allowed for a first-principles description of dielectric,
ferroelectric and piezoelectric properties of arbitrary superlattice
sequences in terms of very few parameters, appropriately arranged
in the form of a cluster expansion.

It is the main scope of this work to extend these ideas to the case
of ferroelectric films with metallic electrodes.
Such an extension is now possible, as there are well-established methods
for treating polarization and electric fields in metal/insulator
heterostructures, and these can readily be combined with
recently-developed approaches for treating the electric
displacement field $D$ as the controlled electric
variable\cite{fixedd,Xifan_sl,nature_mat}.
Using an extensive analysis of several Pt/BaTiO$_3$/Pt and
Au/BaZrO$_3$/Au capacitor heterostructures to illustrate the power
of this approach, we show that a film-electrode interface 
behaves analogously to an insulator-insulator interface in a ferroelectric
superlattice (assuming that there is no Schottky breakdown), in that
the same ``locality principle''~\cite{Xifan_sl} holds.
This means that the film is in a bulk-like state except for
the two or three oxide monolayers which lie closest to the boundary.
Moreover, all the complexity of interfacial chemical bonding
and electrostatics can be incorporated in a single energy contribution,
which we define as the interface electric equation of state.
Taking advantage of the constrained-$D$ technique, we further show how to
extract in practice (from calculations of compositionally \emph{symmetric}
capacitors) such an interface equation of state, and represent it in terms of a
potential drop which is in general a nonlinear function of the electric
displacement field.
Then, we use this information, together with the bulk equation of state
of the ferroelectric, to predict, with full first-principles accuracy,
the electrical properties of capacitors of arbitrary thickness and
geometry (symmetric or asymmetric).
Finally, we show how useful observables such as
polarization and dielectric, piezoelectric and electrostrictive coefficients
are easily evaluated as a byproduct of the above procedure.

Our results demonstrate the validity of $D$ as a fundamental electrical
variable to study ferroelectric capacitors within an ``imperfect 
screening'' regime. (The appropriateness of such an approach was recently
questioned, although in a slightly different context, in Ref.~~\onlinecite{Tagantsev:2008}.)
 From the practical point of view, our detailed study of Au/BaZrO$_3$/Au
capacitors also yields important insight into the similarities and
dissimilarities of AO-terminated versus BO$_2$-terminated perovskite films
in contact with simple-metal electrodes.
On the one hand, the relatively high interfacial capacitances we obtain for both 
interface types corroborate the ideas of Ref.~~\onlinecite{nature_mat}, where 
weak interface bonding was found to be favorable for the overall dielectric 
(or ferroelectric) response of the device.
On the other hand, at the BaO-Au interface we find significant
non-linear effects, which do not fit into a ``constant interfacial 
capacitance'' model. We correlate these effects with the formation
and breaking of the interfacial Au-O bonds upon polarization
reversal. (This same mechanism was already found to strongly influence
the ferroelectric instability in Ref.~~\onlinecite{nature_mat}.)

The manuscript is structured as follows. In Section~\ref{sec_methods} we
review the methodological background and present the new developments
which are specific to this work. In Section~\ref{sec_btopt} we 
discuss the structural and electronic properties of ferroelectric
Pt/BaTiO$_3$/Pt capacitors, which we then use to model their dielectric
and piezoelectric properties as a function of thickness and applied 
bias. In Section~\ref{sec_bzoau} we focus on the Au/BaZrO$_3$/Au model
system. First we compare the electrical properties of the Au-BaO and 
the Au-ZrO$_2$ interface structures; then we show how to reconstruct the 
behavior of asymmetric capacitor configurations starting from the interfacial
and bulk equations of state. Finally, in Section~\ref{sec_discuss} and
Section~\ref{sec_concl} we discuss our results in light of the existing
literature and present our conclusions.

\section{Methods}

\label{sec_methods}

\subsection{Polarization}

\label{secpol}

\subsubsection{Bulk insulators}

We shall consider superlattices and capacitor structures stacked
along $\hat{z}$, so that we are interested in polarizations and
fields only along this direction.  We start with the case of a bulk
insulator, either a single bulk unit cell or a supercell
representing an insulating superlattice, but with a formulation
chosen for convenient later generalization to the case of a
capacitor structure.

We thus consider a periodic insulator described by three real-space lattice
vectors $\mathbf{R}_i$, where for simplicity of notation we impose that
$\mathbf{R}_3=(0,0,c)$ is perpendicular to $\mathbf{R}_{1,2}$
(the latter two lie therefore in the $xy$ plane); the
corresponding reciprocal-space vectors are $\mathbf{G}_{1,2,3}$.
We choose a discrete $k$-point sampling of the form $\mathbf{k}=j\mathbf{b}_\parallel +
\mathbf{k}_\perp$, where the vector
$\mathbf{b}_\parallel= \mathbf{G}_3 / N_\parallel$ spans a
regular one-dimensional mesh of dimension $N_\parallel$, and
$\mathbf{k}_\perp$ belongs to a set of $N_\perp$ special points in the
perpendicular plane.
The electronic ground state is defined by a set of one-particle Bloch
orbitals, $u_{n\mathbf{k}}$; our goal now
is to define the polarization along $\mathbf{G}_3$.

To that end, we first seek a localized representation of the electronic wavefunctions
along the direction $\mathbf{G}_3$ for each given $\mathbf{k}_\perp$.
We do this by constructing a set of maximally localized ``hermaphrodite''
orbitals $w_{n\mathbf{k}_\perp}(\mathbf{r})$ that are Wannier-like along $z$
while remaining Bloch-like in the $xy$ plane
\cite{Sgiarovello/Peressi/Resta:2001,Giustino/Pasquarello:2005}
using a parallel-transport procedure~\cite{Marzari/Vanderbilt:1997}.
The center $z_{n\mathbf{k}_\perp}$ of $w_{n\mathbf{k}_\perp}$ is then defined 
as~\cite{explan-sawtooth}
\begin{equation}
z_{n\mathbf{k}_\perp} = \langle w_{n\mathbf{k}_\perp} | \hat z | w_{n\mathbf{k}_\perp} \rangle =
\int |w_{n\mathbf{k}_\perp}(\mathbf{r})|^2 z dr^3 \,,
\end{equation}
and the contribution of $\mathbf{k}_\perp$ to the polarization is 
\begin{equation}
P (\mathbf{k}_\perp)= \frac{1}{\Omega} \Big( -2e
\sum_n z_{n\mathbf{k}_\perp} + \sum_\alpha Q_\alpha z_\alpha \Big),
\label{eqpins}
\end{equation}
where $z_\alpha$ and $Q_\alpha$ are the ionic coordinate and bare
pseudopotential charge, respectively, of the atom $\alpha$; the
factor of two refers to spin-paired orbitals.
The total polarization $P$ is then obtained by
averaging $\mathbf{k}_\perp$ over its 2D Brillouin zone, while being sure
that the branch choice of $P (\mathbf{k}_\perp)$ is continuous in
$\mathbf{k}_\perp$.

Our Wannier-based definition of $P$, Eq.~(\ref{eqpins}), lends
itself naturally to a local decomposition in terms of the dipolar
contribution of individual oxide layers, as proposed in
Refs.~\onlinecite{Xifan_lp} and \onlinecite{Xifan_sl}.
In particular, given that in typical perovskite insulators the
centers $z_{n{\bf k}_\perp}$ cluster themselves around the oxide layers
they formally ``belong'' to, one can define the layer polarization
(LP) of the $j$-th layer as
\begin{equation}
p_j (\mathbf{k}_\perp)= \frac{e}{S}
\Big( -2\sum_{n \in j} z_{n{\bf k}_\perp} +
\sum_{\alpha \in j} Q_\alpha Z_\alpha \Big).
\end{equation}
In the above equation the sums are restricted to atoms $\alpha$ and Wannier
centers $i$ that are ``in'' the layer $j$, and $S$ is the cell surface area;
again, the overall $p_j$ is calculated by performing a 2D Brillouin zone 
average.
$p_j$ is well defined as long as (i) the oxide layers are charge-neutral,
and (ii) the assignment of
a specific atom or wavefunction to a given layer is clear-cut
and unambiguous.
Both conditions are satisfied in typical II-IV perovskite ferroelectrics such
as PbTiO$_3$ and BaTiO$_3$.

\subsubsection{Capacitor superlattices}

\label{seccapa}

Ideally one would like to study a capacitor in the form of a number
of layers of insulator sandwiched between semi-infinite metallic
contacts.  However, we adopt here the standard
approach of constructing supercells consisting of alternating
insulating and metallic regions stacked along $z$, just as
is normally done when studying ferroelectric superlattices.
We adopt the same notations and conventions as in the previous
subsection, with $c$ being the superlattice repeat distance along $z$.
We set $N_\parallel=1$ (and henceforth write ${\bf k}_\perp={\bf k}$);
this is by no means a limitation, since we are only interested in 
capacitors that are thick enough so that tunneling is insignificant, 
in which case the one-particle bands will have negligible dispersion 
along the $z$ direction.
We further require a rectifying (rather than ohmic) contact at the
oxide/electrode interface. This means that both the valence-band maximum (VBM)
and conduction-band minimum (CBM) of the film are located far enough in
energy from the Fermi level that they are not appreciably populated/depleted
by the tails of the smearing function (e.g., Fermi-Dirac, Gaussian, etc.).

Because the capacitor superlattice is metallic, one might wonder whether
it is possible to define a polarization $P$.
However, the superlattice is only metallic in the $x$ and $y$
directions, whereas we are interested only in computing $P$
along $z$, and only in applying fields along $z$.
The methodology for computing $P$ in such cases was developed in
Ref.~~\onlinecite{Stengel/Spaldin:2007}. 
The electronic states are classified into three energy windows:
\begin{itemize}
\item The completely empty states (upper window) are discarded from the
computation, since they do not contribute to $P$ or to other ground-state
properties.
\item The partially occupied states (middle window) lying in the range
$\mathcal{W}=[E_F-\delta,\Ef+\delta]$
around the Fermi level $\Ef$ are considered as
\emph{conduction} states.
Since these states fall in the energy gap of the dielectric film, they
are confined to the metallic slab, and the dipole moment of their
overall charge distribution is thus well defined.
To make sure that this conduction charge distribution decays fast
enough in the insulating film, it is useful to define its planar average
\begin{equation}
\rho_{\rm cond}(z)= \frac{1}{S}\sum_{\epsilon_{n{\bf k}}\in\mathcal{W}}
w_{\bf k} f_{n{\bf k}} \int dx\, dy\, |\psi_{n{\bf k}}({\bf r})|^2 \, ,
\label{eqrhocond}
\end{equation}
where $\epsilon_{n{\bf k}}$ and $f_{n{\bf k}}$ are the eigenvalue and occupancy of
the state, $w_{\bf k}$ is the $k$-point weight, and $S$ is the cell
cross-sectional area.
\item The lower states, which are all fully occupied, are transformed
to yield a set of hybrid Wannier functions $w_{{\bf k}n}$ that are
maximally localized along $z$ while remaining extended (and labeled by
${\bf k}={\bf k}_\perp$) in the $x$ and $y$ directions.
The contribution of each Wannier function to $P$ is then defined through
the center of the corresponding charge distribution
$\rho_{{\bf k}n}(x)=|w_{{\bf k}n}(x)|^2$.
\end{itemize}
The center of charge of $\rho_{{\rm cond}}(z)$ (middle window) is computed
by integrating against a linear sawtooth function whose discontinuity is
placed in the middle of the insulating region.  Similarly, the center
of each Wannier charge (lower window) is computed from its
$\rho_{{\bf k}n}$ using a sawtooth function whose discontinuity is
chosen far away from its center.
For the systems considered in this work, the $\rho_{{\bf k}n}$ are typically
very well localized, and the main source of error comes from the
slower decay length of $\rho_c$ in the insulator.
This means that in very thin capacitors (few oxide layers) the
polarization becomes ill-defined; rather than a defect of the algorithm,
this is a signature that the system becomes metallic, and the polarization
cannot be defined.

Note that there is an inherent arbitrariness in the separation of
the total charge density into lower and middle windows (i.e., in the choice
of the parameter $\delta$ above).
This arbitrariness indeed affects both $\rho_{{\rm cond}}$ and those $\rho_{{\bf k}n}$ which
lie closest to the electrode; the total value of $P$, however, is not
affected by this choice, and is therefore well defined.
Far enough from the electrode, the $\rho_{{\bf k}n}$ themselves are
unaffected by this choice, and can therefore be used to construct meaningful
layer polarizations, analogously to the case of an insulating superlattice.
This point will be demonstrated in practice in the applications sections.

We are generally concerned with capacitor structures in which a finite
bias is applied across the capacitor.  In our approach, this is treated
by applying a finite macroscopic electric field $\mathcal{E}$ along the
$z$ direction of the superlattice, and identifying $\mathcal{E}c$ as
the bias applied between successive metallic segments.  The formulation
above applies equally well to this case, where it is understood that
the electric field couples to $\rho_{\rm cond}$
(middle window) and to all the Wannier charges (lower window).
Note that the presence of a finite macroscopic field implies that there
is effectively an infinite number of regularly spaced Fermi levels, 
one for each repeated image of the metallic slab along the field direction.
The ``transition'' between two adjacent Fermi levels takes place deep in
the insulating slab, where the system is locally insulating and 
a shift in $E_{\rm f}$ within the gap does not produce any physical 
consequence.

\subsection{Constrained-D method}

\subsubsection{General theory}

We summarize here the details of the constrained displacement-field method
that are most relevant for this work (see Ref.~~\onlinecite{fixedd} for the
full derivation).
For consistency with the previous sections, we restrict our analysis to the
case of a monoclinic system, with the polarization axis, $z$, parallel
to the heterostructure stacking direction and perpendicular to the
$xy$ plane; we shall further assume that $\mathbf{R}_{1,2}$ are
fixed, and only $c$ (together with the ionic and electronic coordinates,
$\{v\}$) is allowed to vary.
Within these assumptions, the constrained-$\mathbf{D}$
method~\cite{fixedd} reduces to a simpler formulation,
where only the $z$ components of the macroscopic fields
$\mathbf{D}$, $\mathbf{P}$ and $\bm{\mathcal{E}}$ are
explicitly treated. Thus, we define the internal energy functional
\begin{equation}
U(D,\{v\},c) = E_{KS}(\{v\},c) + \frac{Sc}{8\pi}
\big[ D - 4\pi P(\{v\},c)\big]^2,
\label{udv}
\end{equation}
which depends directly on the external parameter $D$, and
indirectly on the internal ($\{v\}$) and strain ($c$) variables
through the Kohn-Sham total energy $E_{KS}$ and the
macroscopic polarization $P$; $S=|\mathbf{R}_1 \times \mathbf{R}_2|$
is the constant cell cross-section.
We then proceed to minimize the functional with respect to $v$ and
$c$ at fixed $D$:
\begin{equation}
U(D) = \min_{\{v\},c} U(D,\{v\},c),
\end{equation}
which yields the equilibrium state of the system as a function of the
electric displacement $D$.

$D$ can also be expressed in terms of the reduced variable $d=SD / 4\pi$,
which has the dimension of a charge
and can be interpreted as $d=-Q_{\rm free}$, where $Q_{\rm free}$ is the
free charge per surface unit cell stored at a hypothetical
electrode located at $z=+\infty$~\cite{explan-current}.
Since the surface areas of the parallel plate capacitors
considered in this study are not allowed to vary, constraining
$D$ or $d$ is completely equivalent.
However, for reasons of convenience, we shall adopt $d$ as our
electrical variable in the remainder of this work.

This method is equally valid for bulk insulators, insulating superlattices,
and capacitor superlattices, once the polarization is defined as explained
in Sec.~\ref{secpol}.
For the capacitor case, our adoption of the definitions of
Sec.~\ref{seccapa} implies that the metallic electrode layer
is treated as an infinitely polarizable dielectric, and the free
charges on its surfaces are reinterpreted as polarization charges
coming from the metal.
While such a choice may seem unnatural from the point of view of
textbook electrostatics, it is in fact the most natural one in the
context of first-principles electronic-structure calculations, where
it is not easy to draw a distinction between free and bound charges.
For example, the metal-insulator interface is typically rather diffuse,
with the conduction states of the metal mixing strongly with the
states of the insulator across several interatomic spacings, so that
a spatial distinction is not meaningful, and we have seen in
Sec.~\ref{seccapa} that a distinction based on energy windows is
also arbitrary to some degree.

The reduced electric field $\bar \varepsilon=\mathcal{E}c$, which
is minus the potential step across the
supercell, $\bar \varepsilon= -V$, is related to the internal 
energy by
\begin{equation}
\bar \varepsilon(d) = \frac{dU(d)}{dd}.
\label{eqdudd}
\end{equation}
This corresponds to the fundamental relationship
\begin{equation}
U(D_2) - U(D_1) = \frac{\Omega}{4\pi} \int_{D_1}^{D_2} \mathcal{E}(D) dD.
\end{equation}
of classical electrostatics, but expressed in differential form using
the reduced variables appropriate to the variable-cell case.
The connection to classical electrostatics can be made even more 
apparent by recalling the relationship between the reduced variables and
free charges and potentials
\begin{equation}
U(D_2) - U(D_1) = \int_{d_1}^{d_2} \bar \varepsilon(d) dd = \int_{Q_1}^{Q_2} 
V(Q) dQ.
\end{equation} 

Having established the functional relationships between the active degrees
of freedom (both electrical and structural), it is relatively easy now
to extract from a calculation all functional properties of a material or
device that involve a coupling between them.
For example, the \emph{proper} piezoelectric strain constant can be readily 
obtained as
\begin{equation}
d_{33} = \frac{dc}{d \bar \varepsilon} = \frac{dc}{dd} 
\Big( \frac{d \bar \varepsilon}{dd} \Big)^{-1}.
\label{eqd33}
\end{equation}
Note that in the above equation $d \bar \varepsilon /dd$ has the dimension of an 
inverse capacitance and is related to the free-stress dielectric constant
of the crystal; with the notation of Ref.~~\onlinecite{Wu/Vanderbilt/Hamann:2005} 
we have 
\begin{equation}
\epsilon^{(\sigma)}_{33} = \frac{4 \pi c}{S} \Big( \frac{d\bar \varepsilon}{dd} \Big)^{-1}.
\label{eqeps}
\end{equation}

\subsubsection{Practical procedure}

We typically span the range of relevant polarization states by repeating
the structural and electronic relaxations for a number (five to ten) of
equally spaced $d$ values, $d=d_1,d_2,...,d_n$.
For each $d_i$, we tabulate the energy $U_i$,
(cell-averaged) electric field $\mathcal{E}_i$, and equilibrium
out-of-plane lattice constant $c_i$; we use the latter two
to compute the reduced field $\bar\varepsilon_i = \mathcal{E}_i c_i$.

In order to obtain $\bar\varepsilon_i$ with sufficient accuracy it is important
to perform well-converged structural relaxations, by imposing sufficiently
stringent thresholds on residual forces and stresses.
A useful indicator of the overall numerical quality of the calculations
is the fundamental relationship Eq.~(\ref{eqdudd}), which in principle
should be exactly satisfied.
To check the validity of Eq.~(\ref{eqdudd}) we first perform a
polynomial fit to the calculated $(\bar\varepsilon_i,d_i)$ points. This yields
a continuous function, $\bar\varepsilon(d)$ that we integrate analytically to
obtain $U(d)$ modulo a constant; we choose this constant as the one that best
matches the first-principles internal energies $(U_i,d_i)$.
Any residual discrepancy between $U_i$ and $U(d_i)$ points to a
numerical issue that must be addressed before proceeding further in
the analysis.
Usually the most important source of error concerns the relaxation
of the cell volume and shape; we shall discuss this issue further in
Section~\ref{sec_comput}.

For convenience, we also perform a polynomial fit to the $(c_i,d_i)$
points, which yields a continuous curve $c(d)$ that is relevant for
the piezoelectric response of the crystal, as stated in the
previous subsection.

\subsection{Locality principle and spatial decomposition}

\begin{figure}
\includegraphics[width=2.3in]{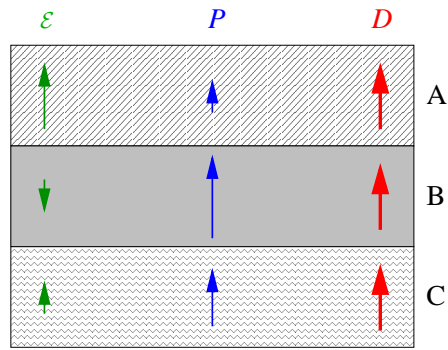} \\
\caption{(Color online) Sktech showing conservation of longitudinal component of
displacement field $D$, but not electric field $\mathcal{E}$ or polarization
$P$, in an insulating superlattice composed of three dielectric constituents
A, B and C.
}
\label{figlocal}
\end{figure}

According to classical electrostatics,
in the absence of free charge the normal component of the electric
displacement field is preserved at a planar interface between two
insulators,
\begin{equation}
({\bf D}_2 - {\bf D}_1)\cdot {\bf \hat{n}} = 0 \, . 
\label{eqdd}
\end{equation}
This means that, for an insulating superlattice in electrostatic
equilibrium, $D$ is the same in all participating layers,
unlike the electric field $\mathcal{E}$ and the polarization $P$, 
whose local values generally vary from layer to layer (see 
Fig.~\ref{figlocal}).
Therefore, using $D$ (or $d$) as the fundamental electrical variable 
is extremely practical for modeling the behavior of ferroelectric
capacitors, because it makes it possible to decompose the equation
of state of a layered structure into the sum of the individual building
blocks.  

For example, we can write the internal energy as
\begin{equation}
U(d) = \sum_i U_i(d),
\end{equation}
where $U_i$ refers to the internal energy of an appropriately defined
sub-unit.
For a capacitor with metallic electrodes, it is natural to
decompose the internal energy as
\begin{equation}
U(d) = \bar U_{\rm L} + U_{\rm L}(d) + N U_{\rm b} (d) + U_{\rm R}(d) + \bar U_{\rm R}
\label{udecomp}
\end{equation}
where $N$ is the number of bulk cells comprising the insulating film
and $U_{\rm b}$ is its bulk internal energy per cell,
$U_{\rm L}(d)$ and $U_{\rm R}(d)$ are the left (L) and right (R) interface
internal energies, and $\bar U_{\rm L}$ and $\bar U_{\rm R}$ are the internal
energies of the left and right metallic electrodes.  (In our
capacitor supercells, $\bar U_{\rm L}$ and $\bar U_{\rm R}$ are combined into
$N_{\rm metal}U_{\rm metal}$, where $N_{\rm metal}$ is the number
of cells of bulk metal and $U_{\rm metal}$ is its internal energy
per cell, which is independent of $d$ as appropriate for a metal.)
Taking the derivative of Eq.~(\ref{udecomp}) according to Eq.~(\ref{eqdudd})
yields
\begin{equation}
\bar\varepsilon(d) = \bar\varepsilon_{\rm L}(d) + 
N \bar\varepsilon_{\rm b} (d) + \bar\varepsilon_{\rm R}(d),
\label{vdecomp}
\end{equation}
where $\bar\varepsilon_{\rm bulk}$ is the potential drop across a unit cell
of the bulk insulator
at a given value of $d$, and $\bar\varepsilon_\mathrm{L,R} = dU_\mathrm{L,R}/dd$
contains the interface-specific information.

The potentials and the energies contain the same information,
apart from a constant of integration, and one can choose to work
with one or the other as a matter of practical convenience.
Indeed, when analyzing the electrical properties of a capacitor,
one is generally interested in energy \emph{differences} between two
different electrical states, rather than in the total energy of
the device.
Therefore, the constant of integration that gets lost in going
from Eq.~(\ref{udecomp}) to (\ref{vdecomp}) is not important for
the scope of our discussion. Thus, we shall assume henceforth
that $U(0)=0$, which also implies that the constant energies
$\bar U_\mathrm{L,R}$ in Eq.~(\ref{udecomp}) have been set to zero.

\subsection{Decomposition of the interface contribution}

\label{secdecomp}

\subsubsection{Partial decomposition}

Since $U(d)$ and $\bar\varepsilon(d)$ in Eqs.~(\ref{udecomp}-\ref{vdecomp})
can be obtained from supercell calculations, while $U_{\rm b}(d)$
and $\bar\varepsilon_{\rm b}(d)$ can be obtained from bulk
insulator calculations, it is straightforward to extract the
quantities
\begin{equation}
U_{\rm int}(d) = U_{\rm L}(d) + U_{\rm R}(d)
\label{eqintU}
\end{equation}
and
\begin{equation}
\bar\varepsilon_{\rm int}(d) = \bar\varepsilon_{\rm L}(d) + \bar\varepsilon_{\rm R}(d)
\label{eqinteps}
\end{equation}
representing the total impact of \emph{both} electrodes on the
electrical equation of state of the capacitor.
Explicitly, we take
\begin{equation}
\bar\varepsilon_{\rm int}(d) = \bar\varepsilon_N(d) - N\bar\varepsilon_{\rm b}(d).
\end{equation}
Often, this is all that is needed, e.g.,
for modeling the polarization and dielectric
response of a given device as a function of the oxide film
thickness (we shall demonstrate this in our first application to
Pt/BaTiO$_3$/Pt capacitors).
The number $N$ of cells of insulating material should
be kept small enough to avoid an undue computational burden, while
remaining large enough to decouple the two electrode interfaces, so that
the center of the oxide slab should behave like the bulk material within
the same mechanical (in-plane strain) and electrical ($d$) boundary
conditions.

\subsubsection{Full decomposition \label{secphi}}

\begin{figure}
\includegraphics[width=3in]{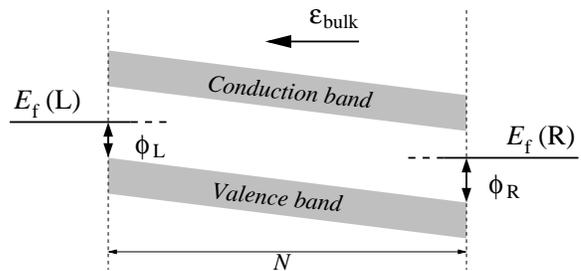}
\caption{ \label{figschottky} Schematic model of the
decomposition of the potential into bulk and interface
contributions. An electron traveling from the left (L)
electrode to the right (R) electrode experiences potential
variations of $-\phi_{\rm L}$, $N\bar\varepsilon_{\rm bulk}$ and
$\phi_R$; the total variation is
$\bar\varepsilon = \Ef({\rm R})-\Ef({\rm L})$.}
\end{figure}

There are, however, situations in which it may be valuable to obtain
the individual terms in Eq.~(\ref{eqinteps}), i.e.,
to define the individual interfacial potential steps
$\bar\varepsilon_{\rm L}$ and $\bar\varepsilon_{\rm R}$ which occur at the left
and right electrode interfaces respectively.
However, instead of using quantities defined as offsets of the average
electrostatic potential across the interface, we find it more physical
to use variables $\phi_{\rm L}$ and $\phi_{\rm R}$ that are the offsets of the
metal Fermi levels $\Ef(L)$ and $\Ef(R)$ relative to the VBM just
inside the insulator, as illustrated in Fig.~\ref{figschottky}.
With this choice, $\phi_{\rm L}$ and $\phi_{\rm R}$ are just the $p$-type
Schottky barrier heights (SBH) at the metal/insulator interface.
(It would be equally viable to adopt the CBM as the reference,
corresponding to $n$-type Schottky barriers, but we do not do so here.)
As long as both electrodes are made from the same material~\cite{explan-electrode},
the total potential step is just $\bar\varepsilon = \Ef(R) - \Ef(L)$.
This difference can be decomposed by following the hypothetical path
in Fig.~\ref{figschottky} of an electron traveling from the left
electrode through the insulator and into the right electrode,
and we obtain
\begin{equation}
\bar\varepsilon = -\phi_{\rm L} + N\bar\varepsilon_{\rm b} + \phi_{\rm R}.
\label{eqphi}
\end{equation}

In the remainder of this section, we make some of the above definitions
more precise, and discuss how in practice to extract
accurate values of the SBH at a polarized metal/insulator interface.
The main issue here is that,
whenever $\bar\varepsilon_{\rm b}$ is non-zero, the SBH is
somewhat ill-defined because the VBM does not have a well-defined
asymptotic value deep in the oxide.  (Instead, it varies linearly
with depth,
with a slope corresponding to the internal electric field
$\mathcal{E}_{\rm b}$).
In the next few paragraphs, we propose a procedure that
provides a reasonable yet sharp definition of $\phi_{\rm L}$ and $\phi_{\rm R}$ even
when $\mathcal{E}_{\rm b} \neq 0$.

\begin{figure}
\includegraphics[width=3.3in]{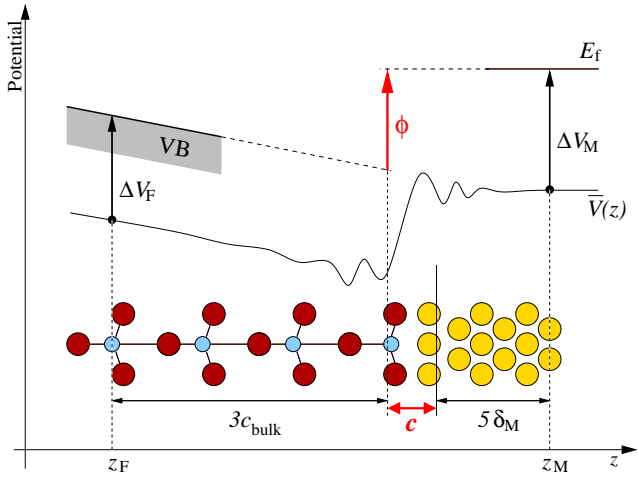}
\caption{(Color online) Illustration of proposed procedure for evaluating the
$p$-type Schottky-barrier height $\phi$ at a perovskite/metal interface
when a macroscopic field is present in the insulating layer.
Light blue circles correspond to B-site cations, red circles
to oxygens, and gold circles to the metal electrode atoms.
}
\label{figsketch}
\end{figure}

For the moment we assume an interface configuration
with the semi-infinite electrode at right and the film
at left.  The situation is sketched in Fig.~\ref{figsketch}, which
also illustrates the following discussion.
We first compute the planar average of the local electrostatic
potential, $V_H(\mathbf{r})$, further convoluted with a
Gaussian filter of width $\alpha$ to suppress the short-range
oscillations,
\begin{equation}
\bar{V}(d,z) = \frac{1}{\sqrt{\pi}\alpha S} \int V_H(d,\mathbf{r}')
e^{-(z-z')^2/\alpha^2} d^3 \mathbf{r}' \;.
\end{equation}
Next, we identify two $z$ coordinates on either side of the
interface, $z_F$ in the film and $z_M$ in the metal.
Both $z_F$ and $z_M$ must be located far enough from the interface that
the short-range structural distortions related to interface
bonding have already relaxed back to the regular bulk-like
spacings of the respective material (oxide film or metal electrode).
As a further requirement, we impose that the interface-related
perturbations in the local electrostatic potential $\bar V(z)$
(schematically indicated in the figure by the strong oscillations
near the metal/film boundary) are also negligible in the neighborhood
of both $z_F$ and $z_M$.
This implies that $\bar V(d,z)$ is a linear function near $z_F$
with a finite slope given by the bulk internal field
$\mathcal{E}_{\rm bulk}(d)$, and $\bar V(d,z)$ is a
constant near $z_M$.
We generally find that three perovskite unit cells on the film
side and five monolayers on the electrode side are sufficient
for both requirements (on the structure and $\bar V(z)$) to be
satisfied accurately.
We therefore set $z_F$ as the $z$ coordinate of the 4-th B-site
cation in the film (numbering as 1 the B cation which lies
adjacent to the interface), and $z_M$ as the
$z$ coordinate of the 6-th metallic layer in the electrode.

Now, using these two reference points in the lattice, we
extract $\bar V(d,z_M)$ and $\bar V(d,z_F)$, which are
indicated in the figure as black circles.
On the film side, we use $\bar V(d,z_F)$ to estimate the VBM at
the interface,
\begin{equation}
E_{\rm VBM}(d) = \bar V(d,z_F) + \Delta V_F(d) + 3\bar\varepsilon_{\rm bulk}(d),
\end{equation}
where $\Delta V_F(d)$ is the relative position of the VBM
to the average electrostatic potential in the bulk (techniques for
calculating this are detailed in
the following subsection).
On the metal side we compute the Fermi energy as
\begin{equation}
\Ef(d) = \bar V(d,z_M) + \Delta V_M,
\end{equation}
where $\Delta V_M$ (independent of $d$) is again a bulk property,
i.e., the relative position of the Fermi level to the average
electrostatic potential of the metal.
Finally, we define the $p$-type Schottky barrier as
\begin{equation}
\phi(d) = \Ef(d) - E_{\rm VBM}(d).
\end{equation}
It is easy to verify that this definition reduces to the standard
technique for calculating Schottky barriers at metal/semiconductor
interfaces~\cite{macroscopic}
whenever the macroscopic field in the oxide vanishes.
Note that the above construction provides, as a byproduct, structural
parameters that are relevant for accessing the piezoelectric
properties of the device.  In particular,
starting from the same $z_F$ and $z_M$, we define an
interfacial expansion
\begin{equation}
c(d) = |z_M - z_F| - N c_{\rm bulk}(d) - 5 \delta_M,
\end{equation}
where $\delta_M$ is the bulk interlayer distance of the metal
(see Fig.~\ref{figsketch}).

Note that, while there is an intrinsic arbitrariness in the definition of
$\phi(d)$ and $c(d)$ (several choices are possible for $z_F$),
the arbitrariness always cancels out in the final equation of state
of the entire capacitor because of the way
these functions are always summed up in pairs (a capacitor
always has two electrodes).
We also note that these functions, by construction, transform properly
under spatial inversion, so that for a capacitor having
a centrosymmetric reference structure, we have
$\phi_R(d) = \phi_L(-d)$ and $c_R(d) = c_L(-d)$.

\subsubsection{$\Delta V_F$ and $\Delta V_M$}

\label{secdeltav}

$\Delta V_M$ can be calculated with high precision for the bulk metal
by extracting the Fermi level and the average electrostatic potential
from the structural and electronic ground state.
To define $\Delta V_F(d)$ we start from a constrained-$D$ calculation of the
bulk oxide. Since a macroscopic electric field is generally present,
the values of both the VBM and the average electrostatic potential are
not directly obvious from the eigenvalue spectrum (strictly speaking,
the energy eigenvalues themselves are ill-defined).
The effect of an electric field is to induce a linear ramp in the electrostatic
potential, and a corresponding linear ``tilting'' of the energy bands.
For a given value of the macroscopic electric displacement, the VBM and the
average electrostatic potential will therefore have the same linear
$z$-dependence, and the difference
$\Delta V_F = V_{\rm VBM}(z) - V_{\rm H}(z)$ will be independent of $z$.
In practice we compute $\Delta V_F$
by first relaxing the structural and electronic degrees of freedom in the
finite field, using the usual convention that the electrons
feel a periodic electrostatic potential having zero unit-cell average,
plus a coupling to the field through the Berry-phase polarization.
$\Delta V_F$ is then obtained by diagonalizing the \emph{zero-field}
Hamiltonian operator in the subspace spanned by the wavefunctions,
which form the ``ground state'' of the finite-field calculation,
and finding its maximum over the wavevectors in the Brillouin zone.

Note that this procedure is to some extent arbitrary, and it is
certainly  possible to adopt alternative strategies. 
Whatever choice is made, the only important requirement is to have
a well-defined reference energy in the insulating lattice as a function
of $D$; the arbitrariness in the specifics of this choice cancel
out anyway when we consider a complete capacitor heterostructure.

\subsection{Computational parameters \label{sec_comput}}

Our calculations are performed within the local-density approximation
of density-functional theory and the projector-augmented-wave
method \cite{Bloechl:1994} as implemented in an ``in-house'' code.
We used a planewave basis cutoff energy of 40 Ry in Section 
\ref{sec_btopt} and of 80 Ry in Section \ref{sec_bzoau}; the 
higher value in the latter case is intended to minimize the Pulay 
error in the stress and the numerical noise in the energies which 
are due to the discrete nature of the basis set~\cite{Froyen/Cohen:1986}.
In all cases we fix the in-plane lattice parameter to a constant
value and we enforce a tetragonal $P4mm$ symmetry constraint;
the out-of-plane lattice parameter, as well as the internal coordinates,
are allowed to relax fully.
The Brillouin zone integrations of the capacitor heterostructures are 
performed with a $6 \times 6 \times 1$ mesh, where $k_z=0$ and the grid 
is shifted in-plane according to the Monkhorst-Pack~\cite{Monkhorst/Pack:1976} 
prescription for two-dimensional sampling; the Gaussian smearing energy is 
set to 0.15\,eV.
In the bulk calculations we use a $6 \times 6 \times 6$ Monkhorst-Pack 
mesh, which is sufficient to converge both the structural and the 
dielectric response of the crystal to an accuracy comparable to that of
the capacitor calculations.
To relax the structure (both internal coordinates and the out-of-plane strain)
at each $d$ value we use a steepest-descent approach, optimally preconditioned
by inverting the force-constant matrix and the elastic constant calculated in the
centrosymmetric $d$=0 configuration.
Generally, five to ten iterations were sufficient to relax the geometries to
a stringent convergence threshold for both forces (10$^{-3}$\,eV/\AA) and 
stresses (10 MPa).
(To ensure excellent accuracy of the calculated energies and potentials,
we further enforce a threshold of 1\% convergence in the internal
electric field.)

Correcting for the Pulay error in the stress is crucial to accurately
model the strain-polarization coupling effects discussed in this work. 
We use a technique similar in spirit to the prescription of 
Ref.~~\onlinecite{Froyen/Cohen:1986}. In particular, we define the corrected 
stress $\sigma_{ij}$ as 
\begin{equation}
\sigma_{ij} = \sigma^0_{ij} + \frac{C \delta_{ij}}{\Omega},
\label{eqsigma}
\end{equation}
where $\sigma^0_{ij}$ is the calculated stress tensor (analytical derivative at 
fixed number of plane waves), $\Omega$ is the cell volume, and $C$ is
a constant (dependent on the cell stoichiometry and plane-wave cut 
off).
To evaluate $C$ several techniques are possible. A possible strategy is 
to fit the dependence of the total energy on the plane-wave cut off,
as discussed in Ref.~~\onlinecite{Froyen/Cohen:1986}.
In our calculations, we infer $C$ by imposing $\sigma_{ij}=0$ in 
Eq.~(\ref{eqsigma}) for a particular configuration of a given system that 
has been structurally relaxed using a different technique (e.g., a Murnaghan 
fit to the energy/volume curve).

\section{Results: {P\lowercase{t}/B\lowercase{a}T\lowercase{i}O$_3$/P\lowercase{t}} capacitors}

\label{sec_btopt}

\subsection{Motivation}

Our goals in this section are threefold. First, we shall introduce our
methods for computing the macroscopic polarization in short-circuited
ferroelectric capacitors, separating the different contributions
that we discussed in Section~\ref{seccapa}.
Second, we shall analyze the structural and electrical properties
as a function of the thickness of the short-circuited film,
identifying those aspects that are common to ferroelectric
single-crystal BaTiO$_3$, and those that depart from the bulk
behavior.
Third, we shall demonstrate with a quantitative model that
even these thickness-dependent perturbations can be understood
in terms of the bulk properties of BaTiO$_3$, once the interface
contribution is properly taken into account.
This analysis is primarily aimed at verifying in practice our
``locality principle,'' which allows us to separate the equation of
state of a capacitor into two interface contributions and a bulk-like
term.
We shall show that, in the case of Pt/BaTiO$_3$/Pt, this separation
works with excellent accuracy down to a thickness of only two
BaTiO$_3$ unit cells.

The choice of Pt and BaTiO$_3$, and more specifically of the 
BaO-terminated interface, is motivated by the recent 
prediction~\cite{nature_mat} of a chemical bonding mechanism
that enhances the ferroelectricity of the film beyond the
bulk BaTiO$_3$ value.
Because of this effect, it was found that Pt/BaTiO$_3$/Pt 
capacitors remain ferroelectric down to a single unit cell 
of BaTiO$_3$, i.e., there is no critical thickness below
which the polar instability is suppressed.
Given the practical interest in overcoming the usually
deleterious size effects in ferroelectric devices, 
the Pt/BaTiO$_3$/Pt system is therefore an appealing test case 
for the present study.
We warn the reader, however, that the above-mentioned features
of the Pt/BaTiO$_3$/Pt system are to some extent anomalous, i.e.,
they depart from the usual understanding of depolarizing effects in 
thin-film ferroelectrics. 
For this reason, the results presented in this section should not be 
understood as an example of the most typical ferroelectric capacitor. 
Instead, this application to Pt/BaTiO$_3$/Pt illustrates how the general 
strategy developed in this work (free from $\it a priori$ assumptions) is 
particularly effective at capturing the peculiar physics of a highly 
non-standard case.
The TiO$_2$-terminated interface of BaTiO$_3$ with Pt would perhaps have
provided a more ``regular'' example, which in principle could have allowed 
us to trace a closer link with earlier first-principles and phenomenological 
results.
However, this system is inappropriate because it suffers from the band-alignment 
issues mentioned in Ref.~~\onlinecite{Junquera_review:2008}. In particular,
we find that the TiO$_2$-terminated BaTiO$_3$/Pt interface has charge-spillage
problems already when the capacitor is in the paraelectric reference structure, 
thwarting attempts at defining a polarization or even introducing an
external bias potential.
In Section~\ref{sec_bzoau} we consider a different ferroelectric/electrode 
combination (BaZrO$_3$/Au) whose BO$_2$-type interface is free from band-alignment 
problems; in that case we are able to compare two different interface types
and discuss the differences between a ``standard'' and a ``non-standard'' case. 

\subsection{Structural and dielectric properties at zero bias}

We start by analyzing the polar ground state
at zero bias of 7, 5, 3, 2 and 1-unit cell thick BaTiO$_3$ films with
compositionally symmetric (the overall spatial symmetry is broken upon
FE off-centering) BaO terminations and Pt electrodes.
($N$-unit cell capacitors are actually
``$N+1/2$'' perovskite cells thick; e.g., $N$=1 means
Pt-BaO-TiO$_2$-BaO-Pt.)
The Pt electrodes are modeled by a 9-layer Pt slab in periodic 
boundary conditions.
We fix the in-plane lattice constant to $a_0=7.276$\,a.u., the theoretical
equilibrium value for cubic SrTiO$_3$, and  we allow
the out-of-plane lattice parameter of the tetragonal supercell, 
as well as the internal coordinates, to relax fully. 
We shall present our results starting from a comparative analysis of 
the relaxed atomic positions in our short-circuited Pt/BaTiO$_3$/Pt 
capacitors; then we shall gradually introduce the ingredients that
enter the definition of the polarization and its coupling to an
external field.

\subsubsection{Structural properties}

\begin{figure}
\begin{center}
\includegraphics[width=3.0in]{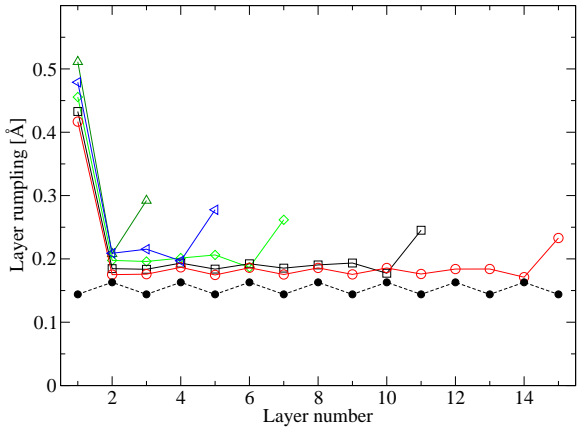} \\
\end{center}
\caption{\label{figrumpl} (Color online)
Layer rumplings, defined as cation displacements relative to oxygens,
for oxide layers in relaxed short-circuited Pt/BaTiO$_3$/Pt capacitors
containing
7 (circles), 5 (squares), 3 (diamonds), 2 (left triangles), or
1 (up triangles) 
perovskite unit cells. Odd and even layer numbers
refer to BaO and TiO$_2$ layers respectively, with Layer 1 being
the BaO layer that is chemically bonded to the Pt.
Bulk values are shown for comparison as filled
symbols connected by dashed lines. The polarization is along $+\hat{z}$.
}
\end{figure}

\begin{figure*}
\begin{center}
\begin{tabular}{c c}
\includegraphics[width=4.0in]{interlayer.eps} &
\hspace{20pt}
\includegraphics[width=1.2in]{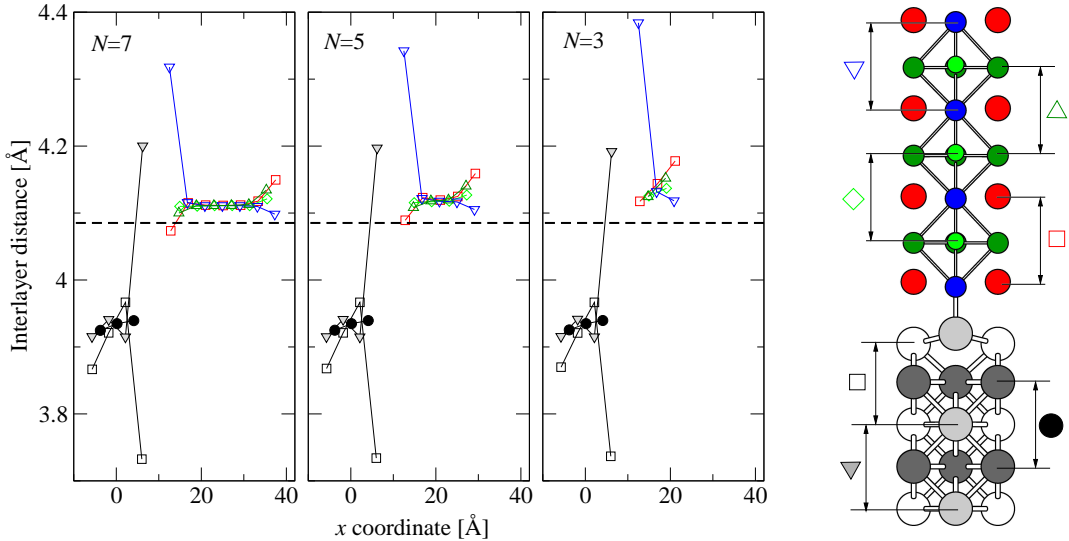}\\
\end{tabular}
\end{center}
\caption{\label{figinterl} (Color online) Relaxed interlayer distances of
short-circuited Pt/BaTiO$_3$/Pt capacitors with oxide thickness of 7, 5 and 3
unit cells and 9 Pt layers.
Black (left) and colored (right) symbols/lines correspond to Pt and
oxide layers respectively.
Vertical axis indicates distances between
neighboring cations belonging to the same sublattice (see right panel);
horizontal axis is the mid-point coordinate.
Dashed line indicates the calculated equilibrium
out-of-plane lattice parameter of bulk BaTiO$_3$ strained to the STO 
in-plane lattice constant. The code for the symbols and colors is
schematically explained in the right panel, where the relaxed structure 
of the bottom interface (located at $x\simeq 10$ \AA)
is shown. Each color corresponds to a different sublattice:
Pt1 (light grey), Pt2 (white), Pt3 (dark grey), Ba (red), Ti 
(light green), O1 (blue), and O2 (dark green).}
\end{figure*}

We plot in Fig.~\ref{figrumpl} our calculated results for the
layer rumplings of the relaxed capacitors at zero bias,
defined as the cation
displacement relative to the oxygens in the same oxide layer;
in the same figure we report the calculated rumpling values for
bulk BaTiO$_3$ as a comparison.
The most striking feature at all thicknesses is
the strong bucking of layer 1, which is the BaO layer directly
in contact with the Pt surface on the negatively polarized
end of the films.
(The BaO buckling is also enhanced at the positively polarized
end, but the effect there is significantly smaller.)
Quite interestingly, the rumplings of all the films are systematically
larger than the bulk values; furthermore, the enhancement in the
structural distortions becomes more important in thinner films.
This is unexpected, as the depolarizing effect is known to
suppress polarization and symmetry-breaking distortions in the
ultrathin limit.
The mechanism leading to such an enhancement is related to the
interfacial chemical bonding effect discussed in Ref.~~\onlinecite{nature_mat};
we shall clarify this point in the following.

In Fig.~\ref{figinterl} we plot the interlayer distances
between ions belonging to the same sublattice, focusing
here on the 3, 5 and 7-cell thick capacitors only.
(Atoms are grouped in different sublattices according to their 
chemical identity. Pt and O atoms are further split into
three and two sublattices, respectively, as shown schematically
in the right panel of Fig.~\ref{figinterl}.)
The atomic layers closest to the interface undergo strong
distortions, both on the electrode and on the insulator side.
The largest effects are located again on the negatively polarized end
of the film. Here, the surface Pt atoms and the BaO ions strongly buckle,
with the overall effect of reducing the Pt-O distance (which ranges from
2.01 \AA{} to 2.04 \AA{} in the capacitors considered) and increasing the
Ba-Pt distance ($\sim 2.9$ \AA).
These features are consistent with the oxygen binding chemically
to the Pt surface, while the Ba atom repels the Pt atom that
lies directly underneath.
Such a picture was proposed, from an analysis of the centrosymmetric
reference structure, in Ref.~~\onlinecite{nature_mat}; here we can see its
impact on the properties of the fully polarized state of the film.
At the positively polarized end of the film, the structural distortions
of the Pt surface are relatively minor, and the oxide film does not
appear to be chemically interacting with the electrode; the Pt-O
distances in all capacitors are larger than 3.3 \AA{}, and the metal-oxide
bonding appears to be of purely electrostatic nature.
Two monolayers away from the interface, the interlayer
distances of the BTO film converge to a uniform value,
which can be understood as the relaxed strain state of
the film in the capacitor heterostructure. In all cases
this value is larger than in the equilibrium value of the
strained bulk, which is indicated in the same figure as a
dashed horizontal line.
(The bulk out-of-plane strain was calculated by imposing
the same in-plane strain as in the capacitor calculations;
therefore, the effect shown in Fig.~\ref{figinterl} is
\emph{not} of mechanical origin.)
Remarkably, the tetragonality of the film increases for thinner
capacitors. Since ferroelectrics have a strong coupling between
polarization and strain, this provides additional evidence to the 
enhancement of polarity we already pointed out earlier while discussing the
layer rumplings.

Note that such a strong coupling makes the results very sensitive to the
accuracy in the relaxation of the out-of-plane lattice constant.
To this end, it is crucial to properly take into account the
effect of the Pulay stress, as explained in Section~\ref{sec_comput}.
In order to check that this procedure was effective, we monitored
in all capacitors the interlayer distance in the center of the Pt
slab, i.e., the black circle at $x=0$ in the three panels of
Fig.~\ref{figinterl}.
The maximum deviation in this value was less than 10$^{-3}$ \AA{},
confirming that our structural relaxations are very well converged.

In the next subsection we shall investigate the electrical properties
of the Pt/BTO/Pt capacitors, and demonstrate the ferroelectric nature of the
enhanced structural distortions discussed above.

\subsubsection{Polarization and electrical properties}

Our goal now is to evaluate the macroscopic polarization of the
capacitor heterostructures discussed in the previous section.
Since all the capacitors are relaxed within standard
short-circuit electrical boundary conditions, the macroscopic
electric field is zero and the polarization is equal to
the electric displacement field $D$.

First we assess the level of accuracy we can expect
for the value of the polarization computed according to
the technique discussed in the methods section.
To that end, we analyze the planar-averaged conduction
charge $\rho_{{\rm cond}}(x)$, as defined in Eq.~(\ref{eqrhocond})
by setting the width of the middle energy window $\delta =0.5$
eV.
Given the Gaussian smearing of $\sigma=0.15$\,eV,
$[\Ef-\delta,\Ef+\delta]$ encompasses all partially occupied states
within an occupancy threshold of 10$^{-6}$. In other words,
at the bottom of the window the smearing function $f(\Ef-\delta)$
is equal to $1-10^{-6}$, while at the top $f(\Ef+\delta)=10^{-6}$.
Recall that all states lying higher in energy are discarded; all
states lying below this window are treated as fully occupied and
transformed into Wannier functions, as we shall discuss shortly.

To define the dipole moment of $\rho_{{\rm cond}}(x)$ it is essential that this
function be small in the middle of the insulating region.
Whenever the film is not thick enough for $\rho_{{\rm cond}}(x)$ to decay to zero,
this introduces an error in the definition of the macroscopic $P$ which can
be roughly estimated by
\begin{equation}
\Delta P \sim  \rho_{{\rm cond}}(L/2) \delta x.
\end{equation}
Here $L/2$ corresponds to the center of the insulating film, and
$\delta x$ is a length that takes into account the arbitrariness in
the positioning of the discontinuity of the sawtooth function used to
define the dipole moment of $\rho_{{\rm cond}}$. We shall assume $\delta x$ to
be half an oxide layer, or 2 bohr units.

\begin{figure}
\begin{center}
\includegraphics[width=2.8in]{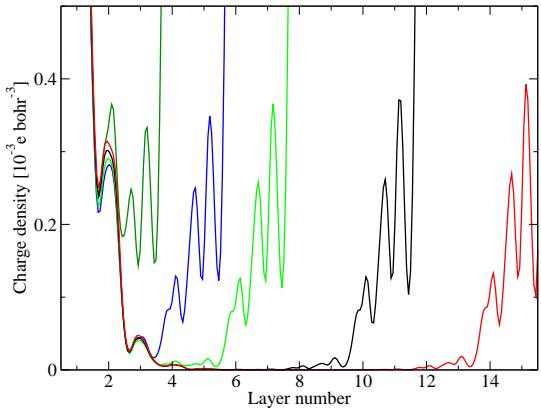} \\
\end{center}
\caption{\label{figcond} (Color online) Conduction charge density
$\rho_{{\rm cond}}(x)$ (Eq.~\ref{eqrhocond}) for five Pt/BaTiO$_3$/Pt capacitors
of Fig.~\ref{figrumpl}.}
\end{figure}

To give an idea of how the conduction charge decays in the oxide films,
we plot in Fig.~\ref{figcond} the calculated $\rho_{{\rm cond}}(x)$ in the five
structures considered.
While the thinnest capacitor (1-cell thick, dark green curve) is clearly
metallic, in all other structures the conduction charge decays almost to
zero in the central oxide layer.
The estimated accuracy using the above formula is of the order of 0.5\,$\mu$C/cm$^2$
for the 2-cell thick capacitor, decreases by roughly one order of magnitude
in the 3-cell thick capacitor, and is essentially zero in the 5- and 7-cell
structures.
In the remainder of this section, therefore, we shall drop the 1-cell structure
from our discussion and focus on the remaining four structures, where
the value of the macroscopic $P$ can be accurately defined.

\begin{figure}
\begin{center}
\includegraphics[width=2.8in]{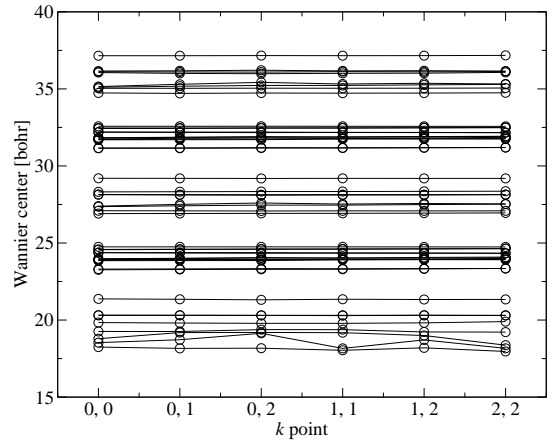} \\
\end{center}
\caption{\label{figwan} Wannier centers in the 2-unit-cell
BaTiO$_3$ film as a function of in-plane $k$-point in the
irreducible 2D Brillouin zone, labeled as $(i,j)$ according to
${\bf k}_\parallel = (i+1/2,j+1/2) / 6$.
Second and fourth groups of centers correspond to TiO$_2$ layers; others
are BaO, of which the first and fifth are in contact with Pt. $P$ points up.}
\end{figure}

The fully occupied states are transformed, separately for each $k$-point,
by means of the parallel transport algorithm~\cite{Marzari/Vanderbilt:1997}.
This yields a set of orthonormal orbitals which sum up to the same charge
density and are maximally localized along the polarization direction (their
Bloch-like character is preserved in plane).
Note that the actual number of states differs for each $k$-point. 
Therefore, when performing the Brillouin zone averages of the polarization,
particular care must be taken in order not to introduce by mistake a fraction
of the quantum of polarization into the final value.
In order to ensure that this issue is properly taken care of, it is useful to
analyze how the Wannier centers distribute in space for each $k$-point.
Upon visual inspection, we find that the Wannier centers are characterized
by a significant degree of disorder in the metallic region, as might be
expected by recalling that the band structure of the metallic slab is not
constituted by full energy bands, and in our algorithm it is abruptly ``cut''
at $\Ef-\delta$.
Conversely, in the insulating region, we find that the Wannier centers cluster
nicely around the oxide layers, analogously to what happens in purely insulating 
superlattices~\cite{Xifan_lp}; to demonstrate such a behavior we plot in 
Fig.~\ref{figwan} the calculated Wannier centers for the 2-cell capacitor.
The total number of orbitals shown in Fig.~\ref{figwan} for each $k$-point
matches exactly the ``nominal'' number of valence orbitals of the oxide ions.
This means that the oxide film can be identified as a charge-neutral
and spatially confined subsystem, whose dipole moment can be computed
with high accuracy, and potential issues with the quantum of polarization
are therefore completely avoided.
(Note the increased $k$-space dispersion in the Wannier centers associated
with the bottom BaO layer; this is due to the strong perturbation induced by
chemical bonding with Pt.)

\begin{figure}
\begin{center}
\includegraphics[width=2.8in]{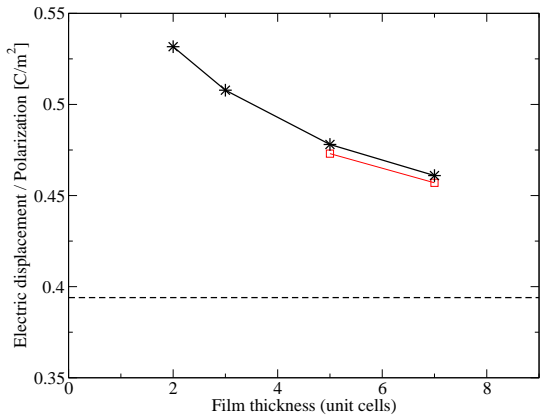} \\
\end{center}
\caption{\label{figbaopol} (Color online) Calculated values of the electric
displacement field (or surface density of free charge stored on the plates)
for short-circuited BaTiO$_3$ capacitors (black solid line and star symbols). The calculated
bulk spontaneous polarization of BaTiO$_3$ at the SrTiO$_3$ in-plane lattice
parameter is shown as a horizontal dashed line. Red squares are the values of the
macroscopic polarization $P_{\rm mac}$ of the 5-unit cell and 7-unit 
cell capacitors, as defined in Eq.~(\ref{eqpmac}).}
\end{figure}

By combining the ingredients discussed in the above paragraphs,
we now compute the electric displacement $D$ of the films,
which is plotted as a function of film thickness in
Fig.~\ref{figbaopol}.
In the thickest 7-layer film $D=46.1$\,$\mu$C/cm$^2$,
which is 17\% larger than the spontaneous polarization of bulk BTO
within the same mechanical boundary conditions (shown as a horizontal
dashed line in the same plot).
This indicates that the electrical boundary conditions induce an
{\it enhancement} in the polarity of the film; this is unlike the vast
majority of cases, where generally a suppression of $P$ due to
depolarizing effects is observed.
The enhancement in $P$ is due to the chemical bonding at the negatively
polarized end of the film to the Pt surface.
Such an effect was discussed for the centrosymmetric geometry in
Ref.~~\onlinecite{nature_mat}; the present study of the fully relaxed capacitors
in short circuit demonstrates that the effect persists in the polar structure.
In fact an analysis of the local electrostatic potential shows that there is,
instead of the usual depolarizing field (which would oppose the spontaneous $P$),
a strong ``polarizing'' field; its magnitude is in excess of 200\,MV/m, and drives
the film significantly more polar than the bulk.

In such a ``negative dead-layer'' regime one would expect thinner films
to be even more polar.
This is nicely confirmed by our results, shown in Fig.~\ref{figbaopol},
where the polarization enhancement attains values as large as 35\%
in the 2-cell thick capacitor; also the tetragonal ratio and the internal
electric fields (not shown) steadily increase with decreasing thickness
as expected from the above qualitative arguments.

\subsubsection{Layer polarizations and macroscopic polarization}

\begin{figure}
\begin{center}
\includegraphics[width=2.8in]{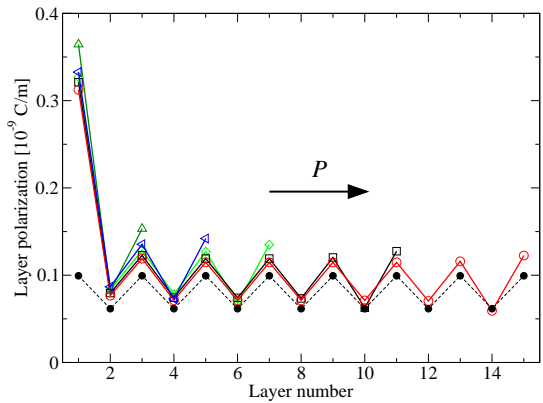} \\
\end{center}
\caption{\label{figlp} (Color online) Calculated Wannier-based layer polarizations for
the capacitors discussed in the text. Even-numbered layers are TiO$_2$;
odd layers are BaO. Bulk values are reported for comparison as filled 
circles connected by dashed lines.}
\end{figure}

While it is not a necessary step to computing the macroscopic $P$,
it is nonetheless interesting to push further the analogy to
insulating superlattices, and compute the layer polarizations (LP)
as defined in Ref.~~\onlinecite{Xifan_lp}.
This involves grouping the Wannier centers of Fig.~\ref{figwan}
into the separate clusters corresponding to the individual oxide
layers, which are obvious from the plot, and computing the
individual dipole moment $p_j$ per surface unit
(see methods section).
We plot in Fig.~\ref{figlp} the results for the Wannier-based layer
polarizations.
In all cases the dipole moment of the first BaO layer is about
three times larger than the LP values in the rest of the films.
This qualitatively reflects the strong structural distortion,
due to the chemical interaction with the Pt surface, which
was discussed in the previous subsection.
Otherwise, the LPs display qualitative features which are
remarkably similar to bulk BaTiO$_3$.
In particular, the LP of both layer types (BaO and TiO$_2$)
have the same positive sign (consistent with the positive
displacement of all cations with respect to the O sublattice),
with the LPs of the BaO layers systematically larger than the
TiO$_2$ values (approximately by a factor of 1.6).
Note that in all cases the LPs are larger than the corresponding
bulk values, and uniformly increase as the film becomes thinner.
This confirms that the enhanced structural distortions
discussed in the previous section correspond indeed to an enhanced
polarization of the films.

It is interesting to note that the LPs converge rather
quickly to a uniform bulk-like sawtooth pattern two or three
oxide layers away from the interface, which indicates that the
perturbations induced by the electrode are rather local.
This contrasts sharply with the picture proposed in a recent
work~\cite{Nardelli} for the same system.
We defer a detailed discussion of this issue to Sec.~\ref{sec_discuss}.

The local LP values in the middle of the ferroelectric,
$p_{\rm BaO}$ and $p_{\rm TiO_2}$, together with the average 
out-of-plane strain $c_{\rm avg}$ inferred from the data of 
Fig.~\ref{figinterl}, provide an accurate estimate of the 
macroscopic $P$ inside the film as
\begin{equation}
P_{\rm film} = \frac{e}{c_{\rm avg}}\big( p_{\rm BaO} + p_{\rm TiO_2} \big).
\label{eqpmac}
\end{equation}
We plot $P_{\rm film}$ of the 5- and 7-unit cell capacitors as square symbols 
in Fig.~\ref{figbaopol}; here $P_{\rm film}$ can be directly compared to the 
calculated values of the electric displacement $D$. 
First, the values of $D$ and $P_{\rm film}$ are very close, as can be 
expected from their relationship (in SI units) 
$D=\epsilon_0 \mathcal{E}_{\rm film} + P_{\rm film}$: 
Indeed, $\epsilon_0 \mathcal{E} << P$ in typical ferroelectrics.
Next, the fact that $D$ is slightly larger than $P$ is consistent with
the electric field's being collinear with $P$ in the capacitors considered 
here, i.e., the interface induces a polarizing effect instead of a 
depolarizing one, as we already discussed extensively.

Note that the electric field $\mathcal{E}_{\rm film}$, as $P_{\rm film}$,
is the macroscopically and planar-averaged value of 
$\mathcal{E}(x)$ inside the film and far from the interfaces.
Therefore, we identify $\mathcal{E}_{\rm film}$ and $P_{\rm film}$ as,
respectively, the \emph{internal field} and the polarization that are 
typically discussed in Landau-Ginzburg models of thin-film ferroelectrics.
As will become clearer in the following sections, the relationship
between $\mathcal{E}_{\rm film}$, $P_{\rm film}$ and $D$ is an
intrinsic property of \emph{bulk} BaTiO$_3$, and does not depend 
on the interfaces, electrical boundary conditions or applied bias
potential. 
This point is crucial to the development of our modeling strategies,
and therefore we consider the above definitions of $P_{\rm film}$ 
and $\mathcal{E}_{\rm film}$ to be very convenient.
Other authors~\cite{Nardelli_apl,Bellaiche_rapid:2005} defined $P$
by averaging the dipoles over the whole volume of the film, including 
the interface region. 
Such a choice is less convenient for modeling, as i) it does not provide
a clear separation between bulk and interface effects; ii) it introduces
a degree of arbitrariness, as the ``bound'' dipoles near the interface are 
strongly mixed with the metallic free carriers.

Summarizing the above results, we have shown that, in thin-film
capacitor configurations with Pt electrodes, BaO-terminated 
BaTiO$_3$ films display many features that are typical of a 
bulk-like crystal within the same mechanical boundary conditions 
(in-plane strain).
However, in addition to these similarities, there are also a few
remarkable departures from the bulk behavior that are fully
consistent with an interface-induced polarization enhancement.
The effect is stronger in thinner films, which is at odds with
the common belief that realistic electrodes would systematically
induce a polarization suppression due to imperfect screening.

In the following we shall use the constrained-$D$ method to understand
the origin of this effect. In particular, we shall demonstrate that
the ``locality principle'' is restored once the long-range electrostatic 
interactions are properly rationalized.
In particular, we shall show that a simple model of this system with
full \emph{ab-initio} accuracy can be constructed in terms of the
electrical properties of bulk BaTiO$_3$ and of the BaO-terminated Pt/BTO
interface.

\subsection{Electrical equation of state}

In order to model the electrical behavior of the Pt/BTO capacitors
considered in this work, we now use the
fixed-$D$ method to sample the equation of state of the 5-cell
thick capacitor; this is the thinnest one in which $\rho_{{\rm cond}}$ is essentially
zero in the middle of the oxide film, which ensures a high level of
accuracy.
For the scope of the present discussion, it is enough to restrict
our investigation to a range of $D$ that encompasses the 
equilibrium values calculated for different thicknesses in short 
circuit, i.e., $0.4 e< d < 0.5e$.
Subsequently, we shall show how this information can be combined with
the bulk equation of state to predict the properties of capacitors of
arbitrary thickness.
We shall start by calculating the electrical equation of state
of bulk BTO.

\subsubsection{Bulk BaTiO$_3$}

\begin{figure*}
\begin{center}
\includegraphics[width=4.6in]{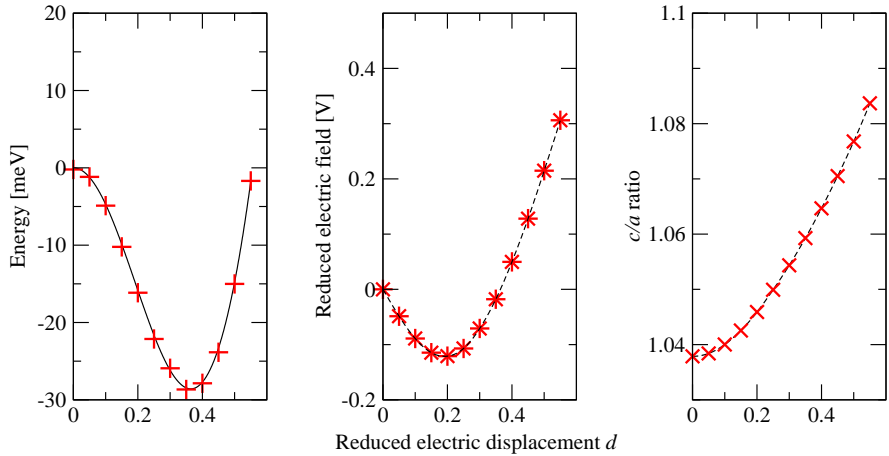}\\ 
\end{center}
\caption{\label{figbulk} (Color online) Internal energy, reduced electric
field, and $c/a$ ratio vs.\ reduced displacement field (in units of $e$)
for coherently strained bulk BaTiO$_3$.
\emph{Ab-initio} data are shown as symbols; dashed curves in the middle and
right panels are spline interpolations; continuous curve in the
left panel is the numerical integral of the spline in the middle panel.}
\end{figure*}

As in the capacitor structures, we fix the in-plane lattice constant to
$a_0=7.276$\,a.u., the theoretical equilibrium value for cubic SrTiO$_3$,
and we let the out-of-plane lattice parameter, as well as the internal
coordinates of the five-atom tetragonal unit cell, relax as a function
of the reduced electric displacement $d$. We use twelve values of $d$,
equally spaced between $d$=0.0 and $d=0.55$.
We extract from the calculation the values of the potential and the
$c$ lattice parameter for every value of $d$; then we use splines
to interpolate these values and we finally integrate the potential
to recover the internal energy $U$. The results are plotted in Fig.~\ref{figbulk}.
Note the perfect match between the values of $U$ calculated 
\emph{ab initio} (orange plus symbols) with the integrated potential 
(green curve); such a good match is a consequence of accurately 
compensating the Pulay stress with a fictitious constant negative 
pressure of $\pi=-2.61$ GPa.
The Pulay error is high (due to the relatively low plane-wave cut off
of 40 Ry), and neglecting it would produce significant errors; with
this simple correction, the numerical values are highly accurate.

The relaxed ferroelectric ground state ($\e$=0)
is at $d_0$=0.364, which corresponds to $P$=39.4\,$\mu$C/cm$^2$, an energy 
$\Delta U$=$-$28.6\,meV/cell, and $c/a=$1.061. 
Note that this is considerably larger than the value, $c/a=$1.038,
in the strained centrosymmetric geometry.
We can understand the $c/a$ ratio of the epitaxially constrained ferroelectric 
phase as a result of two distinct contributing factors. 
One effect comes from the elastic properties (Poisson ratio) of the centrosymmetric
crystal; straining BaTiO$_3$ to the SrTiO$_3$ lattice constant ($-2.1$\%) produces 
a tetragonality of  $c/a=$1.038 even in the absence of a polar distortion.
The second effect, related to polarization-strain coupling, bring this value
to $c/a=$1.061 once the unstable ``soft'' mode is condensed.
For more details about the relationship between polarity and epitaxial strain we
direct the reader to Ref.~~\onlinecite{Ederer/Spaldin:2005}.

\subsubsection{Capacitor structures}
\label{sec:capacstruc}

\begin{figure}[b]
\begin{center}
\includegraphics[width=2.8in]{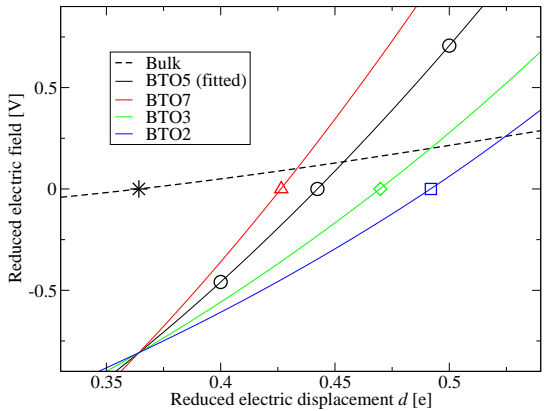} \\
\end{center}
\caption{\label{figpot} (Color online) Calculated values of reduced 
electric field $\e$ as a function of
$d$ in capacitors of various thicknesses (empty symbols). Solid
lines correspond to the first-principles--derived model described
in the text. Dashed line is the calculated
bulk equation of state (from middle panel of Fig.~\ref{figbulk}),
which corresponds to the potential drop across a single bulk unit cell. 
Star corresponds to the relaxed ferroelectric state of the 
strained bulk crystal.}
\end{figure}

We shall write the electrical equation of state of the 5-unit cell
capacitor as $\e_5(d)$, the $d$-dependent reduced electric field.
Since we already have one data point from the calculation
in short circuit ($d$=0.478\,$e$, $\e$=0),
and we expect the potential
to be a rather smooth function of $d$, it is likely that two
additional points lying at the extremes of the interval will be enough
for constructing our model.
Therefore, we repeat the calculation of the 5-cell capacitor twice
with $d$ set to $0.4e$ and $0.5e$, respectively.
(In both cases the ionic positions and out-of-plane lattice parameter
are relaxed to the same convergence thresholds used in the
zero-field cases.)
The smoothness of the potential is confirmed by our results, plotted
as black circles in Fig.~\ref{figpot}, which lie almost on a straight
line.

To account for the small curvature, we interpolate the points with
a second-order polynomial expanded around the spontaneous reduced
displacement $d_0$ of the bulk.  To establish notation, we do this
first for a simple bulk crystal.
Recalling that internal energies are related to the reduced electric 
fields by Eq.~(\ref{eqdudd}), $\e(d) = dU/dd$, and anticipating 
an expansion of the internal energy up to third order in $d-d_0$,
\begin{equation}
U_{\rm b}(d) = A^{\rm b}_0 + A^{\rm b}_1 (d-d_0) +
    \frac{A^{\rm b}_2}{2} (d-d_0)^2 +
    \frac{A^{\rm b}_3}{3!} (d-d_0)^3,
\label{equb}
\end{equation}
we expand $\e_{\rm b}(d)$ as
\begin{equation}
\e_{\rm b}(d) = A^{\rm b}_1 + A^{\rm b}_2 (d-d_0) + \frac{A^{\rm b}_3}{2} 
(d-d_0)^2.
\label{eqvb}
\end{equation}
(Note that the expansion is carried out about the spontaneous
displacement $d_0$ of the bulk, so that $ A^{\rm b}_1$
vanishes by construction.)  We then carry out similar expansions
for each $N$-cell capacitor, e.g., for $N$=5,
\begin{equation}
U_5(d) = A^{(5)}_0 + A^{(5)}_1 (d-d_0) + \frac{A^{(5)}_2}{2} (d-d_0)^2
    +\frac{A^{(5)}_3}{3!} (d-d_0)^3
\label{equ5}
\end{equation}
and
\begin{equation}
\e_5(d) = A^{(5)}_1 + A^{(5)}_2 (d-d_0) + \frac{A^{(5)}_3}{2} (d-d_0)^2.
\label{eqv5}
\end{equation}
The fitted bulk expansion parameters $A_n^{\rm b}$, and the
interface parameters defined via
\begin{equation}
A^{\rm I}_n= A^{(5)}_n -5A^{\rm b}_n ,
\label{Aidef}
\end{equation}
are reported in Table \ref{tab1}.

Focusing first on Eq.~(\ref{eqv5}) for the 5-cell-thick capacitor.
the fitted $\e_5(d)$ is shown as the solid black line passing through 
the circles in Fig.~\ref{figpot}.  Using this together with the
bulk information encoded in Eq.~(\ref{eqvb}), we can then
\emph{predict} the equations of state for thinner
or thicker capacitors according to the formula
\begin{equation}
\e_N(d) = (N-5) \e_{\rm b}(d) + \e_5(d) .
\label{eqpotsum}
\end{equation}
Setting $N$ to 2, 3 and 7 we obtain the colored solid curves in
Fig.~\ref{figpot}.
The intersection of each curve with the $\e=0$ axis yields
a well-defined value of $d$, which is the predicted polarization
state of a capacitor with thickness $N$, and can be directly
compared to the first-principles data already in hand.
To that end, we take the $d$ values from Fig.~\ref{figbaopol} and plot
them as colored symbols on the $\e=0$ axis of Fig.~\ref{figpot}.
The agreement between the first-principles points and the
model predictions is extraordinarily good (discrepancies
are smaller than 0.1\%).
Surprisingly, this holds true even for the thinnest (2-cell)
capacitor, where one would expect the estimation of $d$ to
be less accurate (see previous sections). Also, apart from
purely technical issues in defining $d$, one might expect the
properties of such a thin layer of oxide to depart somewhat
from what is calculated in thicker capacitors.
The accuracy of our model in this thickness regime is indeed
encouraging, and indicates that our methods for accessing the
interfacial electrical properties are able to predict, with
full first-principles accuracy, the behavior of a wide range of
systems.

\begin{figure}
\begin{center}
\includegraphics[width=3in]{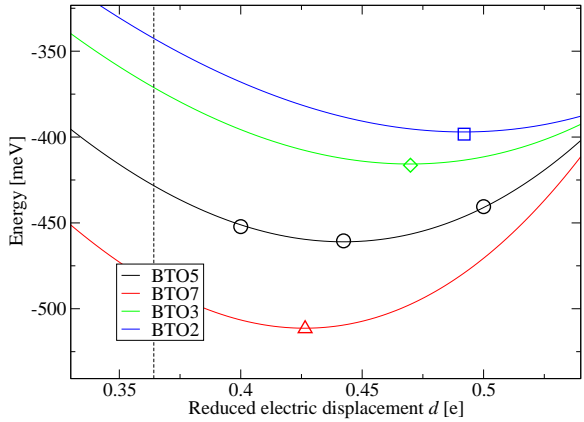} \\
\end{center}
\caption{\label{figene} (Color online) Calculated energies, relative
to corresponding paraelectric state, for BaTiO$_3$ (BTO) capacitors
of various thicknesses and polarization states (symbols).
Solid curves correspond to the model discussed in the text.
Vertical dashed line indicates the spontaneous polarization
of bulk BaTiO$_3$.}
\end{figure}

To further confirm the internal consistency of our
model, we perform a similar analysis for the energetics
of the capacitors.
Let $U_N(d)$ be
the difference in internal energy, for a given thickness $N$, between
the state at specified $d$ and the paraelectric structure at $d$=0.
We plot in Fig.~\ref{figene} the three values of $U_5(d)$ (black
circles) that we extracted for the 5-cell capacitor from the same
calculations described above.
Expanding in $d-d_0$ according to Eq.~(\ref{equ5}), we note that all
the coefficients have already been determined from Eq.~(\ref{eqv5})
except for the arbitrary constant of integration $A^{(5)}_0$.  Adjusting
this one free parameter, we find an excellent match of the fit (black
curve) with all of the data (black circles).
As was done for the potential, we then predict the energy
$U_N(d)$ for other $N$ just by adding or subtracting bulk units,
\begin{equation}
U_N(d) = (N-5)U_{\rm b}(d) + U_5(d).
\label{equsum}
\end{equation}
Again, all of the needed bulk coefficients were already determined from
Eq.~(\ref{eqvb}) except for the constant term $A^{\rm b}_0=\Delta U$.
The results for $N$=2, 3 and 7 are plotted as the colored curves
in Fig.~\ref{figene}.
Again, the minima of all the $U_N(d)$ curves match
very well the points explicitly calculated in the
short-circuit first principles calculations at various
thicknesses.
The maximum discrepancy is about 1\,meV, which is comparable to
the numerical noise in the total energy values introduced by
the discreteness of the plane-wave basis set during variable-cell
structural relaxations.

Two important details are apparent from Fig.~\ref{figpot} and
Fig.~\ref{figene}. First, all curves in Fig.~\ref{figpot} have
a common intersection at $d=d_0$; this is related to the fact that
at $d=d_0$ the internal electric field in the bulk vanishes:
$A^{\rm b}_1=0$ (we shall come back to this point in Section~\ref{sec_band}).
Second, the relaxed internal energies of the capacitors considered here
are about one order of magnitude larger than the depth of the bulk double 
well (see Fig.~\ref{figene}). This indicates that the chemical bonding
mechanism discussed in Ref.~~\onlinecite{nature_mat} has a substantial
impact on the energetics, which explains the strong tendency of the 
capacitors towards a superpolar state.

\subsection{Interfacial dielectric and piezoelectric response}

Our goal now is to show how several useful interface-specific
observables can be directly linked to the {\em electric} equations
of state (EOS) discussed above.
%
%
In addition to the purely electrical variables,
we shall further extend our model by addressing also the {\em elastic}
(i.e., piezoelectric) EOS of both bulk and electrode interface.

\begin{table}
\begin{ruledtabular}
\begin{tabular}{crrrr}
  & \multicolumn{2}{c}{Electrical EOS} &
    \multicolumn{2}{c}{Elastic EOS} \\
  & \multicolumn{1}{c}{Bulk} & \multicolumn{1}{c}{Interface} &
    \multicolumn{1}{c}{Bulk} & \multicolumn{1}{c}{Interface} \\
$n$ & \multicolumn{1}{c}{$A^{\rm b}_n$} &
      \multicolumn{1}{c}{$A^{\rm I}_n$} &
      \multicolumn{1}{c}{$c^{\rm b}_n$} &
      \multicolumn{1}{c}{$c^{\rm I}_n$} \\
\hline
0 & $-$0.0011 & $-$0.0105 & 7.719 & 7.124 \\
1 & 0         & $-$0.0296 & 0.764 & 1.894 \\
2 &    0.0501 &    0.0895 & 0.704 & $-$3.745 \\
3 &    0.0589 &    0.2245 & & \\
\end{tabular}
\caption{ \label{tab1} Values, in atomic units, of the expansion
coefficients used to model the bulk and interface contributions
to the electrical equation of state,
Eqs.~(\ref{equb}-\ref{eqvb}) and (\ref{equi}-\ref{eqvi}),
and to the elastic equation of state,
Eqs.~(\ref{eqcb}) and (\ref{eqci}).}
\end{ruledtabular}
\end{table}

\subsubsection{Dielectric response}
\label{sec:dielresp}

The polynomial expansion of the dielectric response (electrical
EOS) was essentially already determined in
Sec.~\ref{sec:capacstruc}.
Using the $N=5$ as our model structure and using
Eq.~(\ref{eqpotsum}), we single out the interface contribution
by defining the interface EOS to be that of a hypothetical 
``zero-thickness capacitor'',
\begin{equation}
\e_{\rm I}(d) = \e_0(d)=\e_5(d)-5\e_{\rm bulk}(d),
\end{equation}
with a similar relation for the internal energy.  The interface
potential and energy are then expanded as in analogy to
Eqs.~(\ref{equb}-\ref{eqvb}) and Eqs.~(\ref{equ5}-\ref{eqv5}) as
\begin{equation}
U_{\rm I}(d) = \sum_{n=0}^3 \frac{A^{\rm I}_n}{n!} (d-d_0)^n,
\label{equi}
\end{equation}
\begin{equation}
\e_{\rm I}(d) = \sum_{n=0}^n \frac{A^{\rm I}_{n+1}}{n!} (d-d_0)^n .
\label{eqvi}
\end{equation}
The coefficients $A^{\rm I}_n$ are determined once and for all from
a pair of calculations on the bulk and on the 5-cell capacitor
superlattice using Eq.~(\ref{Aidef}).
The resulting bulk and interface coefficients are reported 
in Table \ref{tab1}.

The physical interpretation of the zero-order coefficient
$A^{\mathrm{I}}_{0}$ is immediate, as it represents the
interfacial contribution to the energy of the capacitor when
a reduced displacement $d_0$ is induced in the film (the energy
zero is set to that of the paraelectric $d=0$ state).
The first-order coefficient is also physically transparent,
as it corresponds to the interfacial potential drop at
$d=d_0$.
As we mentioned above, the bulk ground state at $d_0$ has 
zero internal field, so an applied external bias of 
$A^{\rm I}_1$ will always induce the same (bulk-like) polarization, 
regardless of the thickness $N$ of the film.

In order to interpret the higher-order coefficients, we first
derive an expression for the inverse capacitance that is valid for
bulk, interface and the full capacitor structure,
\begin{equation}
C^{-1}_{\rm X}(d) = \frac{d\e^{\rm X}(d)}{dd} =  
A^{\rm X}_2 + A^{\rm X}_3 (d-d_0),
\end{equation}
where X=b,I,$(N)$. 
Therefore, $A^{\rm X}_2$ is the inverse capacitance at $d=d_0$. In the
bulk case we can directly link this coefficient~\cite{fixedd} to the 
static (free-stress) dielectric constant of ferroelectric BaTiO$_3$
through Eq.~(\ref{eqeps}),
\begin{equation}
\epsilon^{(\sigma)}_{33} = \frac{4\pi c}{S A^{\rm b}_2}.
\label{eqeps33}
\end{equation}
Using the values reported in Tab.~\ref{tab1} we obtain 
$\epsilon^{(\sigma)}_{33}=37$.

In the capacitor case the physical meaning of $C^{-1}_{(N)}(d)$ 
is obvious; note that this is an inverse capacitance per surface unit 
cell and must be multiplied by $S$ to obtain an inverse capacitance
\emph{density}.
Finally, $C^{-1}_{\rm I}(d)$ is the inverse interface capacitance that
was discussed, for instance, in Ref.~~\onlinecite{nature_mat}. (Note
that here $C_I$ is the combined effect of \emph{both} interfaces,
unlike the $C_I$ defined in Ref.~~\onlinecite{nature_mat}. We shall
present a general strategy for separating this quantity into two 
individual contributions later, in Section~\ref{sec_bzoau}.)
By using the values of Table~\ref{tab1} we obtain an interfacial 
capacitance density $C_I(d_0) / S$ = 0.44 F/m$^2$;
this value, due to the dielectric non-linearity contained in the
third-order coefficient $A^{\rm I}_3$, is reduced by half 
near the right end of the $d$ interval considered here 
($d \sim 0.55$).

\subsubsection{Piezoelectric response}
\label{sec:piezoresp}

In order to describe piezoelectric effects, we consider the
bulk lattice parameter $c_{\rm b}$ and the interfacial distance $c^I$
as functions of $d$.
In the same spirit as before, we first perform a quadratic fit of
$c_{\rm b}$ as
\begin{equation}
c_{\rm b}(d) = c^{\rm b}_0 +  c^{\rm b}_1(d-d_0) + 
\frac{c^{\rm b}_2}{2}(d-d_0)^2.
\label{eqcb}
\end{equation}
Here $c^{\rm b}_0$ is the equilibrium lattice parameter of the 
epitaxially strained tetragonal state, and $c^{\rm b}_1$ is related
to the piezoelectric constant by Eq.~(\ref{eqd33}), 
\begin{equation} 
d_{33} = \frac{dc_{\rm b}}{d\e} \Big|_{d=d_0} = 
\frac{dc_{\rm b}}{dd} \Big( \frac{d\e}{dd} \Big)^{-1} \Big|_{d=d_0} = 
\frac{c^{\rm b}_1}{A^{\rm b}_2}.
\end{equation}
With the calculated values, we obtain $d_{33}=30$ pm/V. (Note that in
our calculations we kept the in-plane lattice parameters fixed; this
value might change upon full relaxation of the crystal.) 

\begin{figure}
\begin{center}
\includegraphics[width=2.8in]{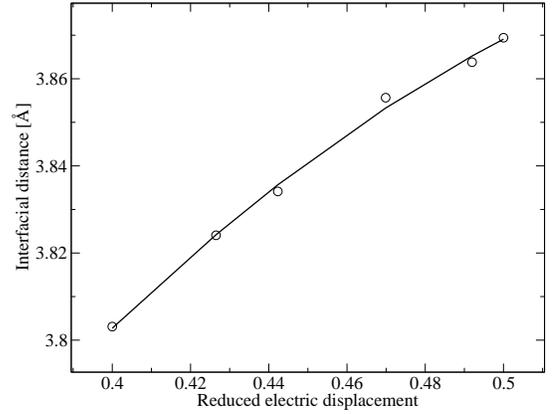} \\
\end{center}
\caption{\label{figpiezo} Calculated interfacial distance $c^I(d)$ of
Eq.~(\ref{eqcintf})
for capacitors of various thicknesses and
polarization states (symbols);
solid curve is a quadratic fit.}
\end{figure}

Next, we define the interface contribution $c^{\rm I}(d)$ as
\begin{equation}
c^{\rm I}(d) = L_N(d) - Nc_{\rm bulk}(d) - n_{\rm Pt} \delta_{\rm Pt}.
\label{eqcintf}
\end{equation}
where $L_N(d)$ is the total thickness of the relaxed
capacitor for a given value of $d$, and $N$ and $n_{\rm Pt}$ are
the number of oxide and metal bulk cells in the capacitor superlattice.
Note that $\delta_{\rm Pt}$, the relaxed interlayer distance in
bulk Pt (strained in plane to the same lattice parameter),
is independent of $d$.
Using this formula, we extracted $c^{\rm I}(d)$ for all the
capacitor structures considered so far, and plotted the values
in Fig.~\ref{figpiezo}.
In the same figure the solid line is a quadratic fit using the
formula
\begin{equation}
c^I(d) = c^I_0 +  c^I_1(d-d_0) + \frac{c^I_2}{2}(d-d_0)^2.
\label{eqci}
\end{equation}
The zeroth-order coefficient $c^I_0$ has the meaning of a combined
effective interface distance for both top and bottom
bottom electrodes.
(It may look larger than expected because there are actually
$N+1/2$ oxide cells in our capacitors, not $N$, and because the
Pt/BTO interface distances are typically somewhat larger than the
bulk interlayer spacings of either Pt or BTO.)
The linear coefficient, in analogy to the bulk case, is
related to the electrode contribution to the piezoresponse
of the capacitor. All the coefficients defined in the text
are reported in Tab.~\ref{tab1}.

In summary, our simple model is able to predict, within the numerical
accuracy of a \emph{full} first-principles calculation, the energy,
polarization, dielectric and piezoelectric response of a Pt/BTO/Pt
capacitor of arbitrary thickness (from two unit cells to infinity).

\subsection{Band lineup and ferroelectricity \label{sec_band}}

The simple model derived above allows us to interpret
the physics behind the polar enhancement
from yet another point of view.
If we go back to Fig.~\ref{figpot}, it is apparent that the
solid lines converge to the same point at a value of $d$
corresponding to the spontaneous polarization of bulk BaTiO$_3$
($d_0=0.364$).
The reason is that the potential is zero in BaTiO$_3$ at
that value of $d$, because the bulk crystal is at electrostatic
equilibrium; hence, adding or subtracting bulk units
does not change the value of $\e_N(d_0)$.
Therefore, $\e_I(d_0) = \e_N(d_0)$, independent of $N$, is an
intrinsic interface property related to the band lineup between
the metal electrode and the fully polarized ferroelectric film,
\begin{equation}
\e_I(d_0) = A^{\rm I}_1 = \phi^+ - \phi^-,
\end{equation}
where $\phi^\pm$ are the respective Schottky barrier heights at
the positively and negatively polarized ends of the film.
Since the symmetry is broken upon ferroelectric off-centering,
these two values generally differ.

Based on typical assumptions of phenomenological theories and on
a large body of experimental data, one would expect a
positive interface potential $\e_I(d_0) > 0$; in short circuit this 
would yield a depolarizing field that \emph{opposes} the polarity of 
the film (recall that $\e=-V$).
This means that generally one needs to apply a positive bias in order to
reach (and sustain) bulk values of $d$ in a thin film;
when the bias is switched off, the polarization is either reduced
or relaxes to zero by transitioning to a multidomain 
state~\cite{Kim_et_al:2005}.
The Pt/BTO/Pt system analyzed in this work displays a rather different
behavior, in that $\e_I(d_0) \sim -0.8$ V is \emph{negative}. This 
is a signature of the strong \emph{polarizing} field discussed earlier
in the context of our calculations in short circuit.

Note that in our calculations so far we never extracted $\phi^+$ and
$\phi^-$ separately, since in the case of compositionally symmetric
capacitors only their difference matters for the electrical properties
of the device.
Analogously, we considered only a \emph{total} interfacial distance
$c_I$ that takes into account both the top and bottom ends of the film.
In the following sections we shall use the techniques described in 
Section~\ref{secphi}
to single out the potential lineup and distances
of either interface individually as a function of $d$.
We shall demonstrate, as an example, how this information allows one to
accurately model the electrical behavior of truly asymmetric devices
starting from calculations performed on systems having a centrosymmetric
paraelectric geometry.

\section{Results: A\lowercase{u}/B\lowercase{a}Z\lowercase{r}O$_3$/A\lowercase{u} capacitors}

\label{sec_bzoau}

\subsection{Motivation}

As a model system for studying the electrical behavior of asymmetric
capacitors we choose ferroelectric devices with Au as the electrode and
BaZrO$_3$ (BZO) as the active film.
This choice is motivated by issues of computational
practicality. As mentioned in the introductory sections,
an essential prerequisite for defining and controlling
the polarization in a metal/insulator heterostructure is its
insulating character along the direction perpendicular to
the interface.
This can only be true if there are non-vanishing Schottky barriers
at both interfaces, and if those barriers are preserved upon
ferroelectric off-centering.
The wider band gap of BZO makes this property much easier to
satisfy than with more conventional ferroelectric oxides such as PbTiO$_3$
and BaTiO$_3$.
Then, given the larger lattice parameter of BZO, we decided
to use Au as the electrode instead of Pt in order to avoid unrealistically
large strains in the electrode slab.
(Pt is more popular in the ferroelectrics community because it matches
better the lattice parameter of many Ti-based perovskites.)
This combination of materials yields large Schottky barriers for both
BO$_2$- and AO- terminated interfaces and is therefore ideally
suited to the scope of the present study.
An interesting aspect of this study is that we are able to compare
the electrical behavior of BO$_2$- and AO-terminated interfaces.

\begin{figure*}
\begin{center}
\includegraphics[width=4.6in]{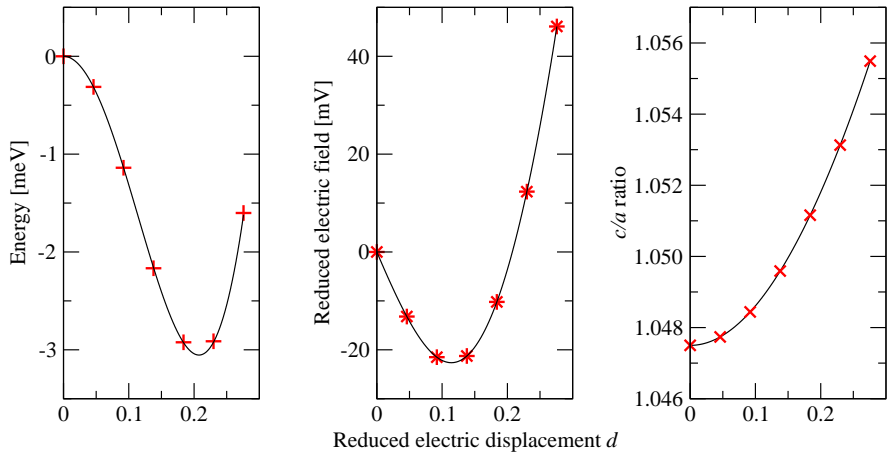}\\
\end{center}
\caption{\label{figbzo} (Color online) Internal energy, reduced electric
field, and $c/a$ ratio vs.\ reduced displacement field (in units of $e$)
for coherently strained bulk BaZrO$_3$.
\emph{Ab-initio} data are shown as symbols; dashed curves in the middle and
right panels are spline interpolations; continuous curve in the
left panel is the numerical integral of the spline in the middle panel.}
\end{figure*}

One important issue is that BZO is not ferroelectric. Experimentally 
it has a stable cubic structure down to very low temperatures, while
theoretically there have been several reports of zone-boundary instabilities 
associated with rotations of the oxygen octahedra~\cite{Bilic_BZO}. 
In order to induce a polar instability in BZO, we set the in-plane
lattice parameter to a fixed value of 7.60\,a.u.\ (a compressive strain of
$-$3.0\% with respect to the theoretical equilibrium lattice parameter of
7.38\,a.u.\ of the cubic structure)~\cite{explan-bazro3}. 
Note that we did not check for possible competing non-polar states, as 
such an analysis would require doubling the size of the simulation cell,
substantially raising the computational cost.
For this reason, our setup should not be understood as a direct
prediction of ferroelectric behavior in epitaxially strained
BaZrO$_3$.
Rather, we intend it primarily as a tractable computational model,
which we expect may be representative of the behavior of a
typical perovskite with a tetragonal ferroelectric ground state (e.g.,
PbTiO$_3$ or BaTiO$_3$ at room temperature).

\subsection{Computational model}

Our model heterostructures consist of (001)-oriented
BZO films with symmetrical ZrO$_2$ or BaO terminations and a thickness 
of $8.5$ unit cells, interfaced with a metal electrode slab of $11$ 
Au monolayers.
Our first goal is to study the full equations of state of these structurally 
\emph{symmetric} capacitors, to establish the similarities and the differences 
arising from the dissimilar (ZrO$_2$/Au vs.\ BaO/Au) bonding configurations.
Next we extract the interface-specific information and use
it to predict the full equation of state of an \emph{asymmetric} 
configuration (with BaO and ZrO$_2$ terminations at opposite ends). 
Then, we verify that our procedure yields the desired result by
comparing this prediction with the
explicitly computed $U(d)$ curve for an asymmetric capacitor having
8 unit cells of BZO and 12 layers of Au.
Before going into details about the capacitor structures, however,
we first briefly summarize the electrical properties of bulk BaZrO$_3$
within the symmetry and mechanical constraints described above.

\subsection{Bulk BaZrO$_3$}

As in the capacitor structures, we fix the in-plane lattice constant to
$a_0=7.60$\,a.u.\ and let the out-of-plane lattice parameter, as well 
as the internal coordinates of the five-atom tetragonal unit cell, 
relax as a function of the reduced electric displacement $d$. We use 
seven evenly spaced values of $d$ ranging between $d$=0.0 and 
$d$=0.276\,$e$.
We extract from the calculation the values of the reduced field
$\bar \varepsilon$ and the
lattice parameter $c$ for each value of $d$, use splines
to interpolate these values, and finally integrate $\bar \varepsilon(d)$
with respect to $d$ to recover the internal energy $U(d)$.
The results are reported in Fig.~\ref{figbzo}.
As for the BaTiO$_3$ case, the match between the values 
of $U$ calculated directly (red `plus' symbols) with those obtained
by integrating  $\bar \varepsilon(d)$ (black curve) is excellent.
Both the spontaneous polarization $P_s$ 
and the double-well potential depth $\Delta U$ are significantly smaller than in
BaTiO$_3$.
The relaxed ferroelectric ground state is at $d$=0.2076, which
corresponds to $P$=20.5\,$\mu$C/cm$^2$, $\Delta U$=$-$3.05\,meV/cell, and
$c/a=$1.052 (compared to $c/a=$1.0475 in the strained centrosymmetric
geometry).

\subsection{Schottky barriers}

\begin{figure}
\begin{center}
\includegraphics[width=3.0in]{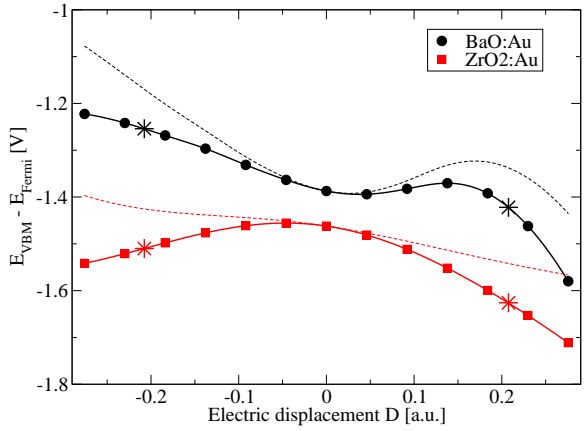} \\
\end{center}
\caption{\label{figphi} (Color online) Solid curves: Interfacial $p$-type Schottky barrier 
values extracted from the symmetric capacitor configurations for BaO-terminated 
(black circles) and ZrO-terminated (red squares) films. Dashed curves: Same,
except that the bulk dependence of the VBM on the electric displacement has been 
subtracted out.  The sign of $d$ follows the convention that the electrode lies
at $z>0$.}
\end{figure}

In Fig.~\ref{figphi} we plot, as a function of the reduced
displacement $d$, the Schottky barrier heights (SBH) extracted
from the symmetric BaO- and ZrO$_2$-terminated capacitors using
the techniques of Sec.~\ref{secphi}.
All values lie between about $-$1.2\,eV and $-$1.7\,eV. Considering that
the calculated LDA gap for centrosymmetric bulk BaZrO$_3$ is 3.12\,eV, 
this indicates that in all cases the Fermi level of Au lies close to 
mid-gap, and that our BZO/Au capacitors are thus free from Schottky-breakdown
issues~\cite{Junquera_review:2008}.
Both curves shown in Fig.~\ref{figphi} have considerable
curvature, although with dissimilar features. 
In the case of ZrO$_2$/Au this curvature is found to be
dominated by the bulk $\Delta V_{\rm F}(d)$: If we remove such a
dependence from $\phi_{\rm ZrO}(d)$, we obtain a roughly linear 
function (red dashed curve, which is related modulo a constant 
shift to the interfacial step in the electrostatic potential).
By contrast, removal of the bulk contribution $\Delta V_{\rm F}(d)$
(black dashed curve) does not restore linearity in the BaO/Au case.

Incidentally, we note that our method may be of general use for 
the computation of Schottky barriers~\cite{Nardelli_prb,Mrovec} 
(independently from their relationship to dielectric properties), 
which are extremely important for many technological applications.
Recall that the SBH has a truly unique definition only
when the macroscopic electric field vanishes in the insulator,
since then one can identify the valence and conduction band
edges precisely.
For non-centrosymmetric insulators such as wurtzite oxides or
spontaneously polarized ferroelectrics, such a condition is not
easily obtained in ordinary first-principles supercell
calculations.
However, our approach makes it extremely easy to do such calculations.
For example, we have indicated with large `star' symbols in Fig.~\ref{figphi}
the ``physical'' values of the SBH, i.e., those corresponding to the
spontaneously polarized ferroelectric film in the absence of any
internal field.

\subsection{Interfacial equation of state: ZrO$_2$/Au}

The discussion in the previous section suggests that there might be 
important qualitative differences between the behavior of BaO/Au and ZrO$_2$/Au
interfaces.
However, it is not immediately obvious how to interpret the $\phi(d)$ 
curves directly, as their relationship to the physical electrical 
response of the capacitor contains some aspects of arbitrariness.
Such arbitrariness does, of course, cancel out when the final equation
of state of the entire capacitor is constructed.
Therefore, in order to obtain quantities that have a direct
physical meaning, we proceed by combining the above $\phi(d)$ curves 
in pairs as appropriate for the capacitor structures of interest.
There are four such structures that we denote as `AB,'
where A and B are variables that specify, for the bottom and top
interfaces respectively, whether the interface is BaO/Au
or ZrO$_2$/Au.
Then the interface contribution to the equation of state of the specified
capacitor structure is
\begin{equation}
\e_{\rm I, AB}(d) = -\phi_{\rm A}(-d) + \phi_{\rm B}(d),
\end{equation}
in terms of which the the equation of state of the entire $N$-cell capacitor
is then $\e_{N,AB}(d) =\e_{\rm I, AB}(d) + N \e_{\rm b}(d)$.

\begin{figure}
\begin{center}
\includegraphics[width=3.0in]{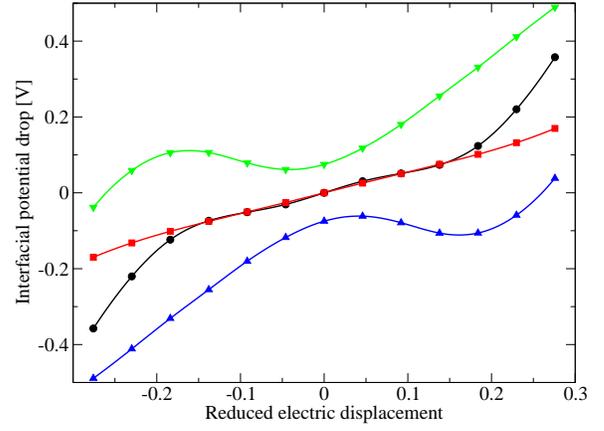} \\
\end{center}
\caption{\label{figvdiff} (Color online) Interfacial equations of state, reconstructed from the
Schottky barrier values of Fig.~\ref{figphi}, for a symmetric BaO- (black circles) or
ZrO$_2$-terminated (red squares) capacitor. The blue upward-oriented triangles refer 
to an asymmetric arrangement with the BaO termination on top; the green downward-oriented
triangles have the BaO termination at the bottom.}
\end{figure}

The four resulting functions $\e_{\rm I, AB}(d)$ are plotted in Fig.~\ref{figvdiff}.
The most striking feature is the almost perfect linearity of the 
symmetric ZrO$_2$/ZrO$_2$ configuration. This means that
the description of the interfacial equation of state in terms
of a constant interfacial capacitance (i.e., replacing the interfaces by
a layer of linear dielectric in series with a bulk-like
BaZrO$_3$ film) is appropriate in this case. The slope of the $\e_{\rm I}(d)$ curve
yields a combined capacitance density of $C_{\rm I} / S$=1.73 F/m$^2$ for 
both interfaces, so that each interface is associated with a capacitance 
density of 3.46 F/m$^2$.
This value is remarkably high when compared to the typical
range of $\sim$0.4-0.6 F/m$^2$ calculated~\cite{nature_2006,nature_mat} 
for oxide electrodes such as SrRuO$_3$.
This result corroborates the ideas proposed in Ref.~~\onlinecite{nature_mat}
that weak electrode-oxide bonding is beneficial to the ferroelectric properties
of a capacitor. Here we indeed find that a chemically inert electrode material
such as Au yields excellent screening and only a marginal perturbation to the
polar response of the film.
Using the formalism developed in Ref.~~\onlinecite{nature_mat}, we find
a ``critical thickness for ferroelectricity'' $N_{\rm crit}=3$ for 
both symmetric geometries (ZrO$_2$/ZrO$_2$ and BaO/BaO). 

\subsection{BaO/Au and bonding properties}

Interestingly, the BaO/BaO curve is almost exactly overlapping with the
ZrO$_2$/ZrO$_2$ one in the interval $-0.15e<d<0.15e$, while a strong
departure from the linear regime occurs for values of $d$ lying outside
this interval.
A significant non-linearity was indeed expected from the $\phi(d)$
curves in Fig.~\ref{figphi}; the fact that this non-linearity cancels for
$-0.15e<d<0.15e$ and yields a quasi-linear behavior is probably 
coincidental.
\begin{figure}
\begin{center}
\includegraphics[width=3.0in]{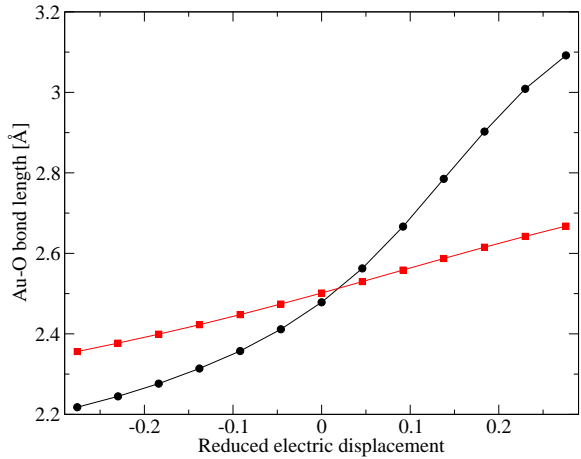} \\
\end{center}
\caption{\label{figbond} (Color online) Interfacial Au-O bond distance as a function 
of electric displacement field.
The sign of $d$ follows the convention that the electrode lies at $z>0$.}
\end{figure}
The non-linearity of the BaO/Au interface emerges most clearly in
the case of an asymmetric capacitor (green and blue curves in
Fig.~\ref{figvdiff}, which are correctly related by a
mirror symmetry operation).

To trace the origin of the qualitative difference between ZrO$_2$/Au
and BaO/Au interfaces (linear vs.\ non-linear behavior),
it is useful to follow the evolution 
of the Au-O bond length as a function of electric displacement,
plotted in Fig.~\ref{figbond}. 
While the bond length varies only weakly and follows a linear trend for
ZrO$_2$/Au, it covers a much wider range of distances (2.2 to 3.1\,\AA)
and displays a strong nonlinearity for BaO/Au.
We interpret the latter behavior as indicative of breaking and
reforming of the Au-O bond upon polarity switching.
Clearly, the breaking of a bond is a highly non-linear event, helping
to explain the calculated features of the electrical response.
This picture agrees fully with the arguments of Ref.~~\onlinecite{nature_mat},
where the bond stability (instability) was correlated with the suppression
(enhancement) of the tendency to polar instability of the capacitor
structure.
The present results corroborate these ideas and provide further evidence
for the strong correlation between interfacial chemistry and
electrical response.

Interestingly, while in the case of Pt/BaTiO$_3$/Pt the interface
bonding mechanism strongly enhances ferroelectricity, in the present case
of Au/BaZrO$_3$/Au the overall effect is a slight suppression.
We shall briefly discuss the origin of this dissimilar behavior in
the following subsection.

\subsection{Enhancement or suppression?}

Why do certain AO-terminated interfaces (e.g., BaTiO$_3$/Pt, BaTiO$_3$/Au) 
enhance the ferroelectric instability of the film, while others 
(especially PbTiO$_3$/Pt but also, to a smaller extent, the  
BaZrO$_3$/Au one discussed here) suppress it instead?
While a definitive answer is not yet available,
some qualitative trends can be explained in terms of the 
frustrated bonding-environment model of Ref.~~\onlinecite{nature_mat}.  
According to this model, a flat AO layer in contact with the electrode produces
a competition between the A-metal repulsion and the O-metal attraction.
The buckling of the AO layer caused by a bulk ferroelectric
distortion of one sign or the other shifts the balance, causing the
bonding or the repulsive force to prevail. 
In fact, even in the centrosymmetric capacitor geometry, the 
interface AO layer is not flat, but exhibits a certain degree of buckling 
(with the A cation typically displacing toward the oxide film)
due to the broken-symmetry environment.
One can therefore expect some difference in behavior between 
perovskite AO-terminated films that show different degrees of ``natural 
buckling'' at the surface of their cubic reference phase.
We computed the values of the AO rumpling of the free PbTiO$_3$,
BaZrO$_3$ and BaTiO$_3$ surfaces, finding values of
0.136, 0.121, and 0.022\,\AA, respectively.
Indeed, these results indicate that the film with the much flatter
surface (BaTiO$_3$) displays a strong enhancement of the polar
instability, while those that are significantly buckled
(PbTiO$_3$ and BaZrO$_3$) do not.
While this is only a rough indication, and other factors are most
likely at work, it suggests a correlation that may help to explain
our detailed numerical results.

\subsection{From symmetric to asymmetric}

\begin{figure}
\begin{center}
\includegraphics[width=3.0in]{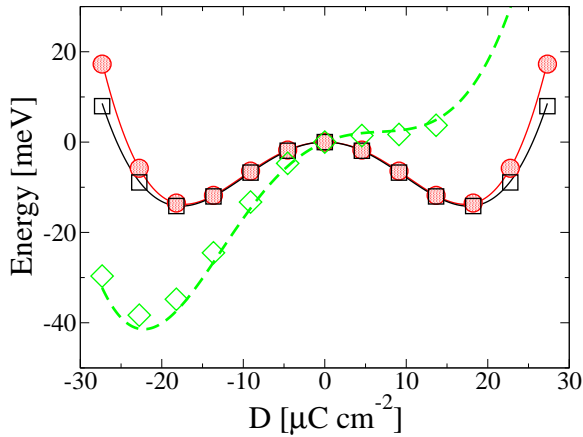} \\
\end{center}
\caption{\label{figasymm} (Color online) Electric equation of state $U(D)$ for
three
capacitor configurations discussed in the text. Curves correspond
to integrating Eq.~(\ref{eqphi}) [or Eq.~(\ref{eqi})], where 
$\phi_\mathrm{L,R}(d)$ were extracted from the symmetric capacitor 
calculations (solid curves in Fig.~\ref{figphi}), and 
$\bar \varepsilon_{\rm bulk}(d)$ was extracted from the bulk calculation
(middle panel of Fig.~\ref{figbzo}). Points correspond to the $U(d)$ 
values explicitly calculated for symmetric BaO:BaO (black squares), 
symmetric ZrO$_2$:ZrO$_2$ (red circles) and asymmetric BaO:ZrO$_2$ (green
diamonds) capacitors.}
\end{figure}

We claimed earlier that it should be possible to use the interface
equations of state extracted from calculations on \emph{symmetric}
capacitors to predict the equation of state for the \emph{asymmetric}
case.  We demonstrate this now.
Our ``asymmetric'' geometry is comprised of an 8-unit-cell
BaZrO$_3$ film and a 12-layer Au slab, where the bottom and
top interfaces (relative to the oxide) are of Au-BaO and ZrO$_2$-Au
type respectively.
Thus, a positive $d$ corresponds to the polarization
pointing towards the ZrO$_2$-Au interface.
The interface-specific contribution to the reduced electric field, 
$\bar \varepsilon_I (d) = -\phi_{\rm L}(d) + \phi_{\rm R}(d)$, 
is plotted as the green curve in Fig.~\ref{figvdiff}.
Note that the curve is linear for $d>0$, while it shows a significant 
non-linearity for $d<0$, where the Pt-O bond breaks, in agreement with 
the discussion of the previous sections.
We now use this function to reconstruct the $\bar \varepsilon (d)$ curve of
the whole capacitor by adding an appropriate number of bulk units as in
Eq.~(\ref{eqphi}),
\begin{equation}
\bar \varepsilon (d) = \bar \varepsilon_I (d) + N \bar \varepsilon_{\rm bulk}
(d) ,
\label{eqi}
\end{equation}
with $N=7.5$~\cite{explan-asymm}. 
Then we numerically integrate $\bar \varepsilon (d)$ to
obtain the $U(d)$ energy curve, which is plotted as the dashed green
curve in Fig.~\ref{figasymm}.

In the same figure we plot two other curves corresponding to the symmetric
geometries, which were obtained from the symmetric curves in
Fig.~\ref{figvdiff}) in the corresponding way (with $N=8$).
For all three curves we also plot, as symbols, the values of $U$ extracted
directly from the first-principles calculations. 
The match is almost perfect for the symmetric cases, as expected since
the potentials and energies were taken from the \emph{same} calculation;
their agreement is just a test of internal consistency.
(Incidentally, note the close agreement between red and black
curves in  Fig.~\ref{figasymm}, especially in the central region,
which is inherited from the similarity of the corresponding red
and black curves in Fig.~\ref{figvdiff}.)
The acid test, however, concerns the asymmetric structure:
the points were extracted from a direct calculation on the
asymmetric structure, while the curve was inferred from the data on
symmetric capacitors using our model of Eq.~(\ref{eqphi}).
The match is excellent, with less than a 1\% discrepancy in the spontaneous 
polarization 
at the minimum in Fig.~\ref{figasymm}, corresponding to short-circuit
boundary conditions.
(The predicted and calculated $D$ values are $-$22.1 and $-$22.2\,$\mu$C/cm$^2$
respectively.)
Note that the spontaneous $D$ is significantly
enhanced ($\sim 8$\%) compared to the bulk value. This is a consequence of the
nonlinearity discussed above, which induces a \emph{positive}
$\bar \varepsilon_I$ for $d<0$ (see green curve in Fig.~\ref{figvdiff}).
The prediction and the actual calculation also nicely agree regarding
the absence of a
secondary minimum at positive $d$. Interestingly, both strategies would yield
such a minimum if the capacitor were just one unit cell thicker; this further
confirms the accuracy of our model.

While all these physical aspects compare very favorably, Fig.~\ref{figasymm}
shows also some discrepancies in the actual values of the internal energy $U$,
in particular concerning the depth of the energy minimum (predicted and
calculated $\Delta U$ values of -$38.4$ and -$41.4$\,meV respectively).
We shall briefly discuss this discrepancy in the following Section.

\subsection{Accuracy issues}

We demonstrated in Section~\ref{sec_btopt}
that the ``locality principle''
established in Section~\ref{sec_methods} holds very accurately, allowing one to predict 
the electrical properties and energetics of capacitors of varying thickness with 
excellent fidelity.
In this respect the discrepancy in the energy curves of Fig.~\ref{figasymm}
is somewhat surprising, and it is worth discussing its
origin to make sure that all aspects of the method are under control.

We find that the cause of the discrepancy is rooted in  
the treatment of the Au electrode in the simulations, rather than in numerical 
(or formal) errors.
In the symmetric calculation, we used an 11-layer Au slab, and from this data we
constructed the green curve in Fig.~\ref{figasymm}. On the other hand, we had to choose
an even (12-layer) Au slab in the asymmetric calculation, which yielded the green
diamond symbols.
It is well known that quantum size effects associated with Fermi-surface nesting
can persist to substantial thicknesses in thin metal films
\cite{Baldereschi/Fall}. 
Thus, it is not unreasonable to expect the surface of a 12-layer slab to behave
slightly differently from that of an 11-layer slab. To prove this point,
we calculated the work function of two free-standing slabs of
11 and 12 layers, and we found a difference of about 60 mV -- enough to 
produce non-negligible discrepancies in the capacitor calculations. 
The use of a finite electronic temperature (to accelerate convergence
with respect to $k$-point sampling) helps, in that it makes the one-particle
density matrix of the metal short-ranged in space.
However, further exploration of these issues falls outside the main scope of the
present work, and we satisfy ourselves with advising the reader of this issue so
that it can be kept in mind when performing future calculations.

\section{Discussion}

\label{sec_discuss}

In the following, we discuss the implications of our work by 
comparing our techniques and results to the relevant literature.

\subsection{Locality}

One of the crucial aspects of our work concerns the use of the
``locality principle'' in the simulation of capacitor structures.
As we mentioned in the introduction, at least two recent theoretical
works have reported phenomena which do not appear consistent with
this assumption. We shall briefly discuss them here.

First, the authors of Ref.~~\onlinecite{Nardelli} reported, for
BaO-terminated Pt/BaTiO$_3$/Pt capacitors (structurally
analogous to those considered in Section~\ref{sec_btopt}),  
a ``ferrielectric'' layer-polarization (LP) pattern, with 
profound qualitative deviations from the bulk pattern, affecting 
the whole volume of the oxide film.
This in principle implies a sharp, qualitative, deviation from the
``locality principle'' mentioned above.
While we cannot provide a definitive explanation for the origin
of the disagreement with our results, we believe it might lie in
the subtleties involved in the LP construction for an overall
metallic system such as the ferroelectric capacitors under
consideration.
In contrast with our work, where localization of the Wannier
states is imposed in one dimension separately for each $k$-point
(as in the original LP formulation in Ref.~~\onlinecite{Xifan_lp}),
the authors of Ref.~~\onlinecite{Nardelli} used fully three-dimensional
maximally-localized Wannier functions.
Furthermore, they treated the metallic states rather differently
than here, in that they adopted a preliminary disentanglement~\cite{Ivo}
procedure before localizing the states by means of the Marzari-Vanderbilt
algorithm~\cite{Marzari/Vanderbilt:1997}.
Finally, the authors of Ref.~~\onlinecite{Nardelli} might have
taken different prescriptions for the assignment of the Wannier functions
to the individual oxide layers, possibly introducing the LP equivalent
of the quantum of polarization in their reported values.
Regardless of which of the above factors may be responsible for
the discrepancy, the effect proposed in Ref.~~\onlinecite{Nardelli}
appears likely to be a consequence of the details of the Wannier
localization/grouping procedure, and we therefore suggest that its
physical significance should be judged with some caution.

Second, the authors of Ref.~~\onlinecite{Duan/Tsymbal:2006} considered
SrRuO$_3$/KNbO$_3$/SrRuO$_3$ capacitors and reported an interface-induced 
disruption of the ferroelectric soft mode of the film,
with the appearance of a head-to-head ``interface domain wall'' located
three unit cells away from the electrode.
This was interpreted as an effect of the strong bonding at the interface,
which would ``clamp'' the interface dipoles to a fixed value; this constraint
would then couple with the ferroelectric instability of the film, producing
the calculated inhomogeneous polarization pattern.
The spatial variation in $P$ is strongly asymmetric, and takes place over 
approximately 6 unit cells at the positively polarized end of the KNbO$_3$
film~\cite{Duan/Tsymbal:2006}. 
This contrasts with our results, where the interface-induced 
distortions are extremely local and heal completely within the first
perovskite unit cell adjacent to the interface.
Even if the metal-insulator interactions are somehow stronger for
the SrRuO$_3$/KNbO$_3$ system than for our cases,
this should just be reflected in a stronger functional dependence
in the interface equation of state, modifying the strength of the
depolarizing field.
One should still expect a uniform polarization deep in the insulator,
unlike
the inhomogeneous polar ground state found by these authors.
Therefore, it is difficult to understand their findings
unless one of the fundamental 
prerequisites for the formalism developed in this work might have
broken down.
For example, Junquera and Ghosez \cite{Junquera_review:2008} have
emphasized the dangers of pathological band alignments which, as an
artifact of the band-gap problem of density-functional
theory, may lead to charge spillage into the perovskite at certain
perovskite-metal interfaces.
We suggest that such a possibility should be investigated
for the SrRuO$_3$/KNbO$_3$ interfaces considered in
Ref.~~\onlinecite{Duan/Tsymbal:2006}.

\subsection{Relationship to Landau theory}

Our approach has many points of contact with earlier Landau-theory
models of depolarization in thin-film ferroelectrics. 
However, there are also some notable differences that we shall
emphasize in the following, in order to avoid confusion or
misunderstanding. 

First, we note that our strategy is different in spirit from
what was done, for example, in Ref.~~\onlinecite{Gerra:2006}. There, the
authors fitted the parameters of a Landau-like expansion
to the calculated first-principles values of the depolarizing 
field in short-circuited capacitors.
In contrast, in our approach the first-principles engine works as a 
stand-alone tool that yields the ground state energy and structure as 
a function of a well-defined electrical variable (within a given set of
specified mechanical/symmetry constraints and thermodynamic ensemble). 
These data can then be fitted \emph{a posteriori} to a polynomial, 
thus obtaining an expression that bears a close resemblance to 
Landau-theory expansions, but the latter is not a necessary step.

Another difference concerns our use of an interfacial capacitance
(or, equivalently, of an effective screening length) that embodies
all the physical ingredients contributing to the electrostatics.
Gerra {\em et al.}~\cite{Tagantsev/Gerra/Setter:2008}, on the contrary, made a 
distinction between purely electrostatic screening and short-range 
chemical bonding effects, and considered them separately in the 
capacitor equation of state.
While such a distinction appears desirable from a conceptual point of
view, implementing it in practice involves a kind of chicken-and-egg problem.
As we have shown in Ref.~~\onlinecite{nature_mat}, screening and interface
bonding are strongly interrelated, and it is not
obvious how to distinguish cause and effect.
At first sight, our approach does not appear to provide a solution to
the above dilemma, as we cast ``everything'' (chemical bonding
and short- and long-range Coulomb interactions) into a black box,
namely, the interface equation of state.
However, on closer inspection our method does implicitly
provide such a separation.
In fact, by explicitly working as a function of a controlled
field (here the $D$ field), we automatically ensure that only 
the ingredients that are electrical in nature are included in 
the interface equations of state, $V_{\rm I}(d)$. 
The non-polar contributions, which are short-ranged and do not have
any direct impact on the electrostatics, merely enter the definition of the 
zero of the energy, and are therefore implicitly (but rigorously) 
singled out.

Finally, we would like to comment briefly on our choice of $D$ as electrical
variable; such a choice is rather convenient, as we have shown in this work. 
Interestingly, using $D$ is frowned upon by some authors in the Landau-theory 
community~\cite{Tagantsev:2008,Bratkovsky_review}, especially in cases where 
depolarizing effects are present. 
A detailed discussion of this issue would bring us far from the main scope of
our work. We limit ourselves to noting that, by means of our 
fixed-$D$ approach, we seek the electronic and structural ground state at 
a given $D$ within a parameter space spanned by all the \emph{microscopic} 
degrees of freedom (which are implicitly present in our energy functional). 
This means that our description fully accounts for the effects of 
the ``background permittivity,'' of hypothetical competing instabilities,
and of electromechanical couplings. For this reason, it is free from the 
shortcomings described in Ref.~~\onlinecite{Tagantsev:2008}.
An important point to stress is that, \emph{in well-behaved cases},
the electrical equation of state of a given system leads to exactly
the same description of the physics
regardless of which independent variable (electric 
field $\mathcal{E}$, polarization $P$, or electric displacement $D$)
is used.
This point is obvious in linear dielectrics, where $P=\chi \mathcal{E}$
and $D=\epsilon \mathcal{E}$. It still holds in non-linear dielectrics
that have a single energy minimum as a function of the applied field 
$\mathcal{E}$.
In more problematic cases, the equation of state might become
multivalued or even singular, depending on the choice of independent 
variable.  In our experience, $D$ tends to be a very convenient
choice, especially in these ``difficult'' situations.

\subsection{Relationship to the effective-Hamiltonian approach}

One of the strengths of our approach is the ability to recast all 
the complexity of the interface interactions into a smooth function 
of a single electrical variable.
This naturally leads to powerful modeling strategies, as we demonstrated  
in practice in Sections~\ref{sec_btopt} and~\ref{sec_bzoau}.
An obvious next step would be to combine the interface
information derived from first principles with higher-level 
``effective Hamiltonian'' descriptions of the ferroelectric 
film~\cite{Zhong/Rabe/Vanderbilt:1994,Zhong/Vanderbilt/Rabe:1995},
in order to describe phenomena that involve larger length and/or 
time scales (e.g., ferroelectric switching).

Effective Hamiltonian approaches have been used quite intensely
in the past few years to investigate size effects in ferroelectric 
nanostructures~\cite{Nishimatsu_et_al:2008,Ghosez/Rabe:2000,
Bellaiche_nature,Bellaiche_prb} such as films, wires and dots.
Generally, the effective Hamiltonian is formulated and fitted
in order to describe \emph{bulk} behavior, as follows.
One first identifies a reduced set of local-mode variables
to describe the amplitudes of the soft ferroelectric mode and
strains within each unit cell.  Then a model of the energy,
written as an expansion in these local-mode degrees of freedom,
is constructed, and the parameters in the expansion are fitted
to a database of first-principles calculations.  Among the
parameters determined in this way are some that correspond to
short- and long-range dipolar interactions. 
The finite-temperature statistical behavior of the system can
then be simulated using Monte Carlo or molecular-dynamics
techniques.

In order to simulate a nanostructure, the effective Hamiltonian
description is typically applied without modification to a
2D, 1D or 0D system.
This approach has allowed for important conceptual advances, 
for instance by elucidating the properties of polarization vortices 
in nanodisks and nanorods;\cite{Bellaiche_nature} in such configurations
the lower dimensionality is largely responsible for the peculiar
behavior.
However, in the case of 2D systems such as ferroelectric
superlattices or thin-film capacitors, an important conclusion 
of our work is that the fine details of the interface bonding
and electrostatics are crucial to determining the overall physical
behavior of the system.
In that sense, the simple abrupt truncation of the dipolar 
interactions~\cite{Bellaiche_rapid:2005} which is assumed in 
the $H_{\rm eff}$ simulations discussed above
may fall short of faithfully reproducing the overall 
response of a realistic device~\cite{nature_mat}.
Including the interface-specific information in a thin-film 
effective Hamiltonian model would therefore be very desirable.

For example, for a given interface, one could evaluate the
interface EOS of the $H_{\rm eff}$ at zero temperature ($T=0$) and 
compare with the one computed ab-initio using the methods described
in this work.
One could then modify the parameters of the $H_{\rm eff}$  in the 
vicinity of the interface until the $T=0$ interface EOS agrees 
with the first-principles one.
This would then enable one
to answer important questions that cannot be directly addressed from
first principles.  For example, what is the temperature dependence of
the interface equation of state? Or, what is the impact of the 
electrode on the stability of the monodomain state versus a 
polydomain one?
Finally, by making use of the ``locality principle'' discussed 
in this work, it would be relatively easy to analyze the $H_{\rm eff}$ 
results and compare them to the fully first-principles values,
with significant benefit for both theories.
To substantiate these arguments, in the following we shall briefly 
discuss two selected $H_{\rm eff}$ works that are particularly
relevant in light of our proposed strategy. 

In Ref.~~\onlinecite{Bellaiche_prb}, Bin-Omram {\it et al.}\ investigate the
impact of electrical and mechanical boundary conditions on the 
polarization and strain of BaTiO$_3$ and Pb(Zr,Ti)O$_3$ (PZT) 
films.
For a BaTiO$_3$ film at a 2.0\% compressive strain, which roughly 
corresponds to the SrTiO$_3$ substrate assumed in our calculations, 
the authors of Ref.~~\onlinecite{Bellaiche_prb} find a spontaneous 
polarization that increases for thinner films, with a value of 
0.56 C/m$^2$ at a thickness of 6 unit cells. This is qualitatively 
similar to the effects we discussed in Section~\ref{sec_btopt} 
for Pt/BaTiO$_3$/Pt capacitors, although significantly 
larger in magnitude.
We stress that such an enhancement is far from being a systematic
property of electroded BaTiO$_3$ films;\cite{nature_mat}
as we suggested above, the surface terms in $H_{\rm eff}$ should
be adapted to the specific electrode interface on a case-by-case basis.

In Ref.~~\onlinecite{Bellaiche/Resta} the authors report that 2D,
1D and 0D ferroelectric nanostructures are characterized by 
``dielectric anomalies'' in the form of a negative internal susceptibility
$\chi^{\rm (int)}$.
It is not unreasonable to think that the surface-induced enhancement of
the ferroelectric instability, which is built into most $H_{\rm eff}$
thin-film models, may be largely responsible for the reported
negative $\chi^{\rm (int)}$ in the 2D case.
Note that $\chi^{\rm (int)}$
is defined there as an average over the volume of the film, unlike our 
definition of the local dielectric permittivity  
$\epsilon(x)$~\cite{explan-bellaiche}, which was introduced in 
Refs.~~\onlinecite{Giustino/Pasquarello:2005} 
and~~\onlinecite{Stengel/Spaldin:2007}.
This further highlights the conceptual advantage of rigorously separating 
bulk and surface/interface effects by performing a \emph{local} analysis, 
rather than global averages.

Overall, we believe that the methods developed in this work open interesting
new avenues for the accurate simulation of ferroelectric nanostructures
within the $H_{\rm eff}$ framework.
 
\subsection{Interpretation of experimental data}

In this section, we ask how one might best make contact between 
an experimental set of electrical measurements on a series of thin-film
capacitors of varying thickness on the one hand, and
the analysis tools we developed in Secs.~\ref{sec_btopt} and
\ref{sec_bzoau} on the other.
Of course, it is best if the experiment can approach
intrinsic conditions insofar as possible.
For example, the experimental film should ideally be in a monodomain state.
This may be hard to achieve in short circuit, as depolarizing effects might 
induce a multidomain state~\cite{Bratkovsky_review, Junquera_domains}, but
often may be obtained by applying a DC bias to the capacitor as in
Ref.~~\onlinecite{Kim_et_al:2005}.
Also, the film should be as free as possible from space charges
arising from charged defects or trapped carriers, which may
contribute to band-bending effects not considered in the theory.

Our first prediction is that there exists a value of the external bias,
$A_1^I$, that yields the same value $d_0$ of the spontaneous electric
displacement regardless of the film thickness
(provided that the interfaces and the film quality are similar).
Our second prediction is that, around this bias value, the inverse 
capacitance of the films should scale linearly with thickness, with the
coefficient of proportionality being directly related to the bulk permittivity.
Assuming a small range of biases around $A_1^I$, we can discard the third-order 
coefficients $A_3$ and write
\begin{equation}
V_N(d) = A_1^I + (d-d_0) A_2^{(N)} 
\label{eqvlin}
\end{equation}
with
\begin{equation}
A_2^{(N)} = A_2^{\rm I} + N A_2^{\rm b} .
\label{eqclin}
\end{equation}
We end up with three coefficients that describe the 
electrical properties of the capacitors and their dependence on thickness.

To obtain this information, we believe it is best to represent the data in 
a $(P,V)$ plot as in Fig.~\ref{figpot} rather than a $(P,\mathcal{E})$ plot 
as is usually done.
In this way, one obtains direct visual insight into the existence of
a common intersection point $(P_0,A_1^I)$.
Note that the bulk $A_2^{\rm b}$ coefficient provides, 
as a byproduct, useful information on the dielectric properties of the film
through Eq.~(\ref{eqeps33}),
as it is independent of the specific interface. (In principle, it depends only 
on the applied strain due to epitaxial matching, and of course on the
operating temperature.)
By combining Eqs.~(\ref{eqvlin}) and (\ref{eqclin}) we obtain, for the
spontaneous polarization of a short-circuited capacitor of a given 
thickness $N$,
\begin{equation}
P_0^{(N)} \sim P_0^{(b)} - \Big( \frac{4\pi}{S} \Big) \frac{A_1^I}{A_2^{\rm I} + 
N A_2^{\rm b}}.
\end{equation}
Note that, in typical phenomenological models, the interface is treated 
as a linear dielectric, which implies $A_1^I = d_0 A_2^I$. $A_2^I$ is 
related to the interface capacitance $C_I$ and to the effective screening 
length $\lambda_{\rm eff}$ by
\begin{equation}
\frac{1}{2}\,A_2^I = C_I^{-1} = \frac{4\pi}{S} \lambda_{\rm eff}.
\end{equation}
The factor of one-half on the left-hand side relates to the assumption
of two equivalent interfaces with identical electrical properties.
 
\section{Conclusions}

\label{sec_concl}

We have developed a comprehensive methodological framework for the computation and
analysis of ferroelectric capacitors with realistic electrodes.
Our method is based on density-functional theory, and on recently-developed techniques
for performing calculations at a given value of the electric displacement field.
By making a rigorous separation between the interface and bulk contributions 
to the electrical equation of state of a capacitor, we obtain a compact model, 
of full first-principles accuracy, for the electrical (and piezoelectric) response
as a function of bias potential and thickness.
We expect these advances to facilitate the comparison of theory with
experimental data.
We also hope that it will stimulate a fruitful interaction with other theoretical 
approaches based, e.g., on Landau theory or effective Hamiltonians.
Application of similar strategies to investigating the interface coupling between
electric polarization, magnetism and other structural degrees of freedom (such as
octahedral tilting) that were not considered here are under way.

\section*{Acknowledgments}

This work was supported by the Department of Energy SciDac program on
Quantum Simulations of Materials and Nanostructures, grant number 
DE-FC02-06ER25794 (N.A.S. and M.S.) and by ONR grant N00014-05-1-0054 
(D.V.).

\bibliography{Max}

\begin{thebibliography}{51}
\bibitem[{exp({\natexlab{a}})}]{explan-sawtooth}
\bibinfo{note}{In practice, we represent the position operator in periodic
  boundary conditions by means of a saw-tooth function, with the discontinuity
  placed in the region where the Wannier charge is smallest in magnitude.}
\bibitem[{exp({\natexlab{b}})}]{explan-current}
\bibinfo{note}{$d$ can be interpreted as the time integral of the \emph{total}
  transient current that has flowed through a facet of the supercell while
  adiabatically polarizing the system. This means that $d$ is inclusive of the
  displacement current, which stems from variations in the macroscopic
  $\mathcal{E}$ even in the absence of a polarizable material.}
\bibitem[{exp({\natexlab{c}})}]{explan-electrode}
\bibinfo{note}{We specialize here to the case in which both electrode materials
  are the same. Otherwise, a term should be included to reflect the different
  separation between the Fermi level and average electrostatic potential in the
  two metals.}
\bibitem[{exp({\natexlab{e}})}]{explan-asymm}
\bibinfo{note}{Since we assumed $N=8$ for the non-stoichiometric symmetric
  geometries, each interface term contains half of the interface oxide
  monolayer, and one must take this into account by setting $N=7.5$ in the
  asymmetric case.}
\end{thebibliography}

\end{document}